\documentclass[11pt,twoside, final]{amsart}
\usepackage{mathrsfs}
\copyrightinfo{0}{Iranian Mathematical Society}
\pagespan{1}{\pageref*{LastPage}}
\usepackage{etoolbox,lastpage}
\commby{}
\date{\scriptsize   Received: , Accepted: .}
\usepackage{latexsym,amsmath,amsthm,amscd,amsfonts,amssymb,enumerate}
\usepackage{graphicx}		
\usepackage{color}
\usepackage[colorlinks]{hyperref}

\theoremstyle{definition}

\theoremstyle{remark}

\numberwithin{equation}{section}
\begin{document}

%% The title of the paper goes here.  Edit your title.

\title[Constacyclic codes over $\mathbb{Z}_{p^s}+u\mathbb{Z}_{p^s}$]{On a class of constacyclic codes over the non-principal ideal ring $\mathbb{Z}_{p^s}+u\mathbb{Z}_{p^s}$}
%% Now edit the following to give First Author name and address:
%% $^*$ for the corresponding author.

\author[Y. Cao]{Yuan Cao}
\address[Yuan Cao]{School of Sciences, Shandong University of
 Technology, Zibo, Shandong 255091, China}
\email{woodwest2@189.cn}

\author[Y. Cao]{Yonglin Cao $^*$}
\address[Yonglin Cao]{School of Sciences, Shandong University of
 Technology, Zibo, Shandong 255091, China}
\email{ylcao@sdut.edu.cn}
%% If there are three of more authors they are added in the obvious way.

  \thanks{$^*$Corresponding author}
%------------------------------------------------------------------------------------%
%%
%% Use the following command to make the title for the paper.
%
 %\CoverPage

 \maketitle
%
%%% The following environment is needed for the abstract.
%%%

\begin{abstract}
$(1+pw)$-constacyclic codes of arbitrary length over the non-principal ideal ring $\mathbb{Z}_{p^s} +u\mathbb{Z}_{p^s}$
are studied, where $p$ is a prime, $w\in \mathbb{Z}_{p^s}^{\times}$ and $s$ an integer satisfying $s\geq 2$. First,
the structure of any $(1+pw)$-constacyclic code over $\mathbb{Z}_{p^s} +u\mathbb{Z}_{p^s}$ are presented. Then enumerations for the number of all codes and the number of codewords in each code, and the structure of dual codes for
these codes are given, respectively. Then self-dual $(1+2w)$-constacyclic codes over $\mathbb{Z}_{2^s} +u\mathbb{Z}_{2^s}$
are investigated, where $w=2^{s-2}-1$ or $2^{s-1}-1$ if $s\geq 3$, and $w=1$ if $s=2$.
\end{abstract}

\section{\bf Introduction}
\par
    Algebraic coding theory deals with the design of error-correcting and error-detecting codes for the reliable transmission
of information across noisy channel.
   The class of constacyclic codes play a very significant role in
the theory of error-correcting codes. The most impotent classes of
these codes are that of cyclic codes and negacyclic codes, which have been well studied
since the late 1950s. Since 1999, special classes of constacyclic
codes over certain classes of finite commutative chain rings have been studied
by numerous authors. See [1-17], for example.

\par
    Let $A$ be a finite commutative ring with identity $1\neq 0$, $A^{\times}$ the multiplicative group of units of
$A$ and $a\in
A$. We denote by $\langle a\rangle_A$ or $\langle a\rangle$ the ideal of $A$ generated by $a$, i.e. $\langle
a\rangle_A=aA$. For any ideal $I$ of $A$, we will identify the
element $a+I$ of the residue class ring $A/I$ with $a$ (mod $I$) in this paper.

   For any positive integer $N$, let
$A^N=\{(a_0,a_1,\ldots,a_{N-1})\mid a_i\in A, \ 0\leq i\leq N-1\}$ which is an $A$-module with componentwise addition and scalar multiplication by elements of $A$. Then an $A$-submodule ${\mathcal C}$ of $A^N$ is called a \textit{linear code} over $A$ of length $N$.
For any vectors $a=(a_0,a_1,\ldots,a_{N-1}), b=(b_0,b_1,\ldots,b_{N-1})\in A^N$.
The usual \textit{Euclidian inner product} of $a$ and $b$ is defined by
$[a,b]_E=\sum_{j=0}^{N-1}a_jb_j\in A$.
Then $[-,-]_E$ is a symmetric and non-degenerate bilinear form on the $A$-module
$A^N$. Let ${\mathcal C}$ be a linear code over $A$ of length $N$. The \textit{Euclidian dual code}
of ${\mathcal C}$ is defined by ${\mathcal C}^{\bot_E}=\{a\in A^N\mid [a,b]_E=0, \ \forall
b\in {\mathcal C}\}$, and ${\mathcal C}$ is said to be \textit{self-dual} if ${\mathcal C}={\mathcal C}^{\bot_E}$.

   Let $\gamma\in A^{\times}$.
Then a linear code
${\mathcal C}$ over $A$ of length $N$ is
called a $\gamma$-\textit{constacyclic code}
if $(\gamma a_{N-1},a_0,a_1,\ldots,a_{N-2})\in {\mathcal C}$ for all
$(a_0,a_1,\ldots,a_{N-1})\in{\mathcal C}$. Particularly, ${\mathcal C}$ is
a \textit{negacyclic code} if $\gamma=-1$, and ${\mathcal C}$ is
a  \textit{cyclic code} if $\gamma=1$.
  For any $a=(a_0,a_1,\ldots,a_{N-1})\in A^N$, let
$a(x)=a_0+a_1x+\ldots+a_{N-1}x^{N-1}\in A[x]/\langle x^N-\gamma\rangle$. We will identify $a$ with $a(x)$ in
this paper. By [11] Propositions 2.2 and 2.3, we have the following conclusions.

\vskip 3mm \noindent
  {\bf Lemma 1.1}  \textit{Let $\gamma\in A^{\times}$. Then ${\mathcal C}$ is a  $\gamma$-constacyclic code
of length $N$ over $A$ if and only if ${\mathcal C}$ is an ideal of
the residue class ring $A[x]/\langle x^N-\gamma\rangle$}.

\vskip 3mm \noindent
  {\bf Lemma 1.2}  \textit{The dual code of a $\gamma$-constacyclic code of length $N$ over
$A$ is a $\gamma^{-1}$-constacyclic code of length $N$ over
$A$, i.e., an ideal of $A[x]/\langle
x^N-\gamma^{-1}\rangle$}.

\vskip 3mm \par
   Recent years, codes over finite non-principal ideal commutative rings have been studied by many authors.
For example, in Yildiz et al [18],
MacWilliams identities, projections, and formally self-dual codes for linear codes over the ring $\mathbb{Z}_4+u\mathbb{Z}_4$ and their application to  real and complex lattices have been studied.

\par
   From now on, let $p$ be an arbitrary prime, $s$ an integer satisfying $s\geq 2$ and
$\mathbb{Z}_{p^s}=\mathbb{Z}/\langle p^s\rangle=\{0,1,2,\ldots,p^s-1\}$. Denote $\mathbb{Z}_{p^s}[u]/\langle u^2\rangle$ by
$\mathbb{Z}_{p^s}+u\mathbb{Z}_{p^s}=\{a+ub\mid a,b\in \mathbb{Z}_{p^s}\}$ ($u^2=0$), which is a non-principal ideal ring.
The operations on
$\mathbb{Z}_{p^s}+u\mathbb{Z}_{p^s}$ are defined by:
$$\alpha+\beta=(a+b)+u(c+d) \ {\rm and} \ \alpha\beta=ac+u(ad+bc),$$
for any $\alpha=a+bu,\beta=c+du\in \mathbb{Z}_{p^s}+u\mathbb{Z}_{p^s}$ with $a,b,c,d\in \mathbb{Z}_{p^s}$.
For any fixed $w\in\mathbb{Z}_{p^s}^{\times}$,
 the following questions
have not been investigated completely for $(1+pw)$-constacyclic codes over $\mathbb{Z}_{p^s}+u\mathbb{Z}_{p^s}$ of arbitrary length
to the best of our knowledge:

\par
  (Q-1) Present  precisely all distinct $(1+pw)$-constacyclic codes over $\mathbb{Z}_{p^s}+u\mathbb{Z}_{p^s}$ of arbitrary length $N$, and count the number of these codes.

\par
  (Q-2) For each code ${\mathcal C}$ presented above, determine the number of codewords contained in ${\mathcal C}$ and
give the dual code of ${\mathcal C}$ precisely.

\par
  (Q-3) Determine the self-duality for $(1+pw)$-constacyclic codes over $\mathbb{Z}_{p^s}+u\mathbb{Z}_{p^s}$.

\par
    The present paper is organized as follows. In Section 2, we investigate the structure and properties of
the ring $\mathbb{Z}_{p^s}[x]/\langle x^{p^kn}-(1+pw)\rangle$.
In Section 3, we give a canonical form decomposition for any $(1+pw)$-constacyclic code  of length $p^kn$ over $\mathbb{Z}_{p^s}+u\mathbb{Z}_{p^s}$
$(u^2=0)$, list all distinct codes by their generator sets and enumerate the number of all codes and the number
of codewords in each code respectively. By use of the canonical form decomposition, we
obtain the dual code of each code and investigate the self-duality of these codes in Section 4. In Section 5, we
list all 6419 self-dual $3$-constacyclic code over $\mathbb{Z}_8+u\mathbb{Z}_8$
of length $14$.

%%%%%%%%%%%%%%%%%%%%%%%%%%%%%%%%%%%%%%%%%%%%%%%%%%%%%%%%%%%%%%%%%%

%%%
%%% Section 2
%%%
\section{\bf Direct Sum Decomposition of The Ring $\mathbb{Z}_{p^s}[x]/\langle x^{p^kn}-(1+pw)\rangle$}

\par
 From now on, let $N=p^kn$ where $p$ is a prime, $k,n$ are positive integers satisfying
${\rm gcd}(p,n)=1$. We consider how to decompose the ring $\mathbb{Z}_{p^s}[x]/\langle x^{N}-(1+pw)\rangle$
into a direct sum of finite chain rings where $s\geq 2$ and $w\in \mathbb{Z}_{p^s}^{\times}$. This decomposition will be used in the following
sections.

\vskip 3mm \noindent
  {\bf Lemma 2.1} ([10] Proposition 2.1) \textit{Let $A$ be a finite associative and commutative
  ring with identity. Then the following conditions are equivalent}:

\par
   (i) \textit{$A$ is a local ring and the maximal ideal $M$ of $A$ is principal, i.e., $M=\langle \pi\rangle$
for some $\pi\in A$};

\par
   (ii) \textit{$A$ is a local principal ideal ring};

\par
   (iii) \textit{$A$ is a chain ring with ideals $\langle \pi^i\rangle$, $0\leq i\leq \nu$,
where $\nu$ is the nilpotency index of $\pi$}.

\vskip 3mm \noindent
  {\bf Lemma 2.2} ([10] Proposition 2.2) \textit{Let $A$ be a finite commutative chain ring,
with maximal ideal $M=\langle\pi\rangle$, and let $\nu$ be the
nilpotency index of $\pi$. Then}
\par
   (i) \textit{For some prime $p$ and positive integer $m$, $|A/\langle \pi\rangle|=q$ where $q=p^m$,
$|A|=q^{\nu}$, and the characteristic of $A/\langle \pi\rangle$ and
$A$ are powers of $p$};
\par
  (ii) \textit{For $i=0,1,\ldots,\nu$, $|\langle\pi^i\rangle|=q^{\nu-i}$}.

\vskip 3mm\noindent
    {\bf Lemma 2.3} ([14] Lemma 2.4)\textit{Using the notations in Lemma 2.2, let
$V\subseteq A$ be a system of representatives for the equivalence classes of $A$ under
congruence modulo $\pi$. (Equivalently, we can define $V$ to be a maximal subset of $A$ with the property that
$r_1-r_2\not\in \langle \pi\rangle$ for all $r_1,r_2\in V$, $r_1\neq r_2$.) Then}

\vskip 2mm \par
  (i) \textit{Every element $a$ of $A$ has a unique $\pi$-adic expansion: $a=\sum_{j=0}^{\nu-1}r_j\pi^j$,
$r_0,r_1,\ldots$, $r_{\nu-1}\in V$}.

\vskip 2mm \par
  (ii) \textit{$|A/\langle \pi\rangle|=|V|$ and $|\langle \pi^i\rangle|=|V|^{\nu-i}$ for
$0\leq i\leq \nu-1$}.

\vskip 3mm\par
   Let $a\in\mathbb{Z}_{p^s}$. Then $a$ has a unique $p$-adic expansion:
\begin{center}
$a=\sum_{j=0}^{s-1}p^ja_j, \ a_0,a_1,\ldots,a_{s-1}\in \mathbb{F}_p=\{0,1,\ldots,p-1\},$
\end{center}
where we regard $\mathbb{F}_p$ as a subset of $\mathbb{Z}_{p^s}$ (but $\mathbb{F}_p$ is not a subring of $\mathbb{Z}_{p^s}$). It is well known that $a\in \mathbb{Z}_{p^s}^{\times}$
if and only if $a_0\neq 0$.
Denote $\overline{a}=a_0\in \mathbb{F}_p$. Then
$^{-}: a\mapsto \overline{a}$ ($\forall a\in \mathbb{Z}_{p^s}$) is a ring homomorphism from
$\mathbb{Z}_{p^s}$ onto $\mathbb{F}_p$, and this homomorphism can be extended to a ring homomorphism from
$\mathbb{Z}_{p^s}[y]$ onto $\mathbb{F}_p[y]$ by $\overline{f}(y)=\sum_{i=0}^m\overline{b}_iy^i$
for any $f(y)=\sum_{i=0}^mb_iy^i\in \mathbb{Z}_{p^s}[y]$.
A monic polynomial $f(y)\in \mathbb{Z}_{p^s}[y]$
of positive degree is said to be \textit{basic irreducible} if $\overline{f}(y)$ is an irreducible
polynomial in $\mathbb{F}_p[y]$ ([16] Chapter 13, Page 328).

\vskip 3mm \noindent
   {\bf Lemma 2.4} \textit{Let $f(x)$ be a monic basic irreducible polynomial in $\mathbb{Z}_{p^s}[x]$ of
degree $d$ and denote $\Gamma=\mathbb{Z}_{p^s}[x]/\langle f(x)\rangle$. Then}

\par
  (i)  ([16] Theorem 14.1]) \textit{$\Gamma$ is a Galois ring of characteristic $p^s$ and cardinality $p^{sd}$. Moreover,
$\Gamma=\mathbb{Z}_{p^s}[\zeta]$ where $\zeta=x+\langle f(x)\rangle\in \Gamma$ satisfying
$\zeta^{p^{d}-1}=1$}.

\par
   \textit{Denote $\overline{\Gamma}=\mathbb{F}_{p}[x]/\langle \overline{f}(x)\rangle$ and $\overline{\zeta}=x+\langle \overline{f}(x)\rangle\in \overline{\Gamma}$. Then $\overline{\Gamma}=\mathbb{F}_p[\overline{\zeta}]$ which is a finite field of cardinality $p^d$,
$\overline{f}(x)=\prod_{i=0}^{d-1}(x-\overline{\zeta}^{p^i})$ and that $^{-}$ can be extended to a ring homomorphism from
$\Gamma$ onto $\overline{\Gamma}$ by $\xi\mapsto \overline{\xi}=\sum_{j=0}^{d-1}\overline{a}_j\overline{\zeta}^j$,
for all $\xi=\sum_{j=0}^{d-1}a_j\zeta^j\in \Gamma$ where $a_0,a_1,\ldots,a_{d-1}\in \mathbb{Z}_{p^s}$}.

\par
  (ii) ([5] Lemma 2.3(ii)) \textit{$f(x)=\prod_{i=0}^{d-1}(x-\zeta^{p^i})$}.

\vskip 3mm\noindent
   {\bf Theorem 2.5}  \textit{Let $f(x)$ be a monic basic irreducible polynomial in $\mathbb{Z}_{p^s}[x]$ of
degree $d$, $w_0\in \mathbb{Z}_{p^s}^{\times}$, denote ${\mathcal R}=\mathbb{Z}_{p^s}[x]/\langle f(x^{p^k}(1+pw_0)^{-1})\rangle$ and set}
\begin{center}
${\mathcal T}=\{\sum_{j=0}^{d-1}a_jx^j\mid a_0,a_1,\ldots,a_{d-1}\in\mathbb{F}_p\}\subseteq {\mathcal R}$
\end{center}
\textit{in which we regard $\mathbb{F}_p$ as a subset of $\mathbb{Z}_{p^s}$. Then}

\vskip 2mm\par
   (i) \textit{${\mathcal R}$ is a finite chain ring with maximal ideal $\langle f(x)\rangle$
generated by $f(x)$, i.e., $\langle f(x)\rangle=f(x){\mathcal R}$, the nilpotency
index of $f(x)$ is equal to $p^ks$ and ${\mathcal R}/\langle f(x)\rangle$ is a finite field of cardinality $p^{d}$}.

\vskip 2mm\par
   (ii) \textit{Each $\alpha\in {\mathcal R}$ has a unique $f(x)$-adic expansion:
$\alpha=\sum_{j=0}^{p^ks-1}b_j(x)f(x)^j$, $b_0(x),b_1(x),\ldots,b_{p^ks-1}(x)\in {\mathcal T}.$}

\vskip 2mm\par
   (iii) \textit{For each integer $l$, $1\leq l\leq p^ks$, let $\langle f(x)^l\rangle$ be the
ideal of ${\mathcal R}$ generated by $f(x)^l$. Then
$\langle f(x)^l\rangle=\{\sum_{j=l}^{p^ks-1}b_j(x)f(x)^j\mid b_l(x),\ldots,b_{p^ks-1}(x)\in {\mathcal T}\},$
and every element $\beta$ of the residue class ring ${\mathcal R}/\langle f(x)^l\rangle$ has a unique $f(x)$-adic expansion:
$\beta=\sum_{j=0}^{l-1}b_j(x)f(x)^j, \ b_0(x),\ldots,b_{l-1}(x)\in {\mathcal T}$.}

\par
\textit{Moreover, $\beta$ is an invertible element of ${\mathcal R}/\langle f(x)^l\rangle$,
i.e. $\beta\in ({\mathcal R}/\langle f(x)^l\rangle)^{\times}$,
if and only if $b_0(x)\neq 0$. Hence $|\langle f(x)^l\rangle|=p^{d(p^ks-l)}$ and $|({\mathcal R}/\langle f(x)^l\rangle)^{\times}|=(p^{d}-1)p^{(l-1)d}$}.

\vskip 3mm\noindent
   \textit{Proof} Let ${\mathcal R}=\mathbb{Z}_{p^s}[x]/\langle f(x^{p^k}(1+pw_0)^{-1})\rangle$. It is known that
\begin{equation}
f(x)^{p^k}=p\vartheta(x) \ {\rm in} \ {\mathcal R}, \ {\rm where} \  \vartheta(x)\in {\mathcal R}^{\times}
\end{equation}
(A direct proof for this equation is given in Appendix). Then
\begin{equation}
\langle p\rangle=\langle f(x)^{p^k}\rangle\subseteq \langle f(x)\rangle
\end{equation}
as ideals of ${\mathcal R}$, and hence $f(x)^{p^ks}=p^s\vartheta(x)^s=0$ in ${\mathcal R}$ by Equation (2.1).

\par
   (i) Let $J=\langle p,f(x)\rangle$ be the ideal of ${\mathcal R}$ generated by
$p$ and $f(x)$. Then
\begin{eqnarray*}
{\mathcal R}/J&=&(\mathbb{F}_p[x]/\langle \overline{f(x^{p^k}(1+pw_0)^{-1})}\rangle)/\langle \overline{f}(x)\rangle
= (\mathbb{F}_p[x]/\langle \overline{f}(x)^{p^k}\rangle)/\langle \overline{f}(x)\rangle\\
&=&\mathbb{F}_p[x]/\langle\overline{f}(x)\rangle,
\end{eqnarray*}
up to natural ring isomorphisms, where $\mathbb{F}_p[x]/\overline{f}(x)\rangle$ is a finite field of $p^d$ elements by Lemma 2.4(i). Hence $J$ is a maximal ideal of ${\mathcal R}$. Since
$f(x)^{p^ks}=0$, both $p$ and $f(x)$ are nilpotent elements of ${\mathcal R}$. From this one can verify easily
that every element in ${\mathcal R}\setminus J$ is invertible, which implies that ${\mathcal R}$ is a local ring with $J$ as its unique
maximal ideal. Furthermore, by Equation (2.2) we conclude that $J=\langle f(x)\rangle$ and so
${\mathcal R}/\langle f(x)\rangle=\mathbb{F}_p[x]/\langle\overline{f}(x)\rangle$.

\par
   As stated above, by Lemma 2.1 we see that ${\mathcal R}$ is a finite chain ring with the unique maximal ideal
generated by $f(x)$. Let $z$ be the nilpotency
index of $f(x)$. By Lemma 2.2(i) it follows that $|{\mathcal R}|=|\mathbb{F}_p[x]/\overline{f}(x)\rangle|^z=p^{z d}$.
On the other hand, by ${\deg}(f(x^{p^k}(1+pw_0)^{-1}))=p^kd$ it follows that
\begin{center}
$|{\mathcal R}|=|\mathbb{Z}_{p^s}[x]/\langle f(x^{p^k}(1+pw_0)^{-1})\rangle|=(p^s)^{p^kd}=p^{p^ksd}$.
\end{center}
Therefore, $z=p^ks$.

\par
  (ii) By ${\mathcal R}/\langle f(x)\rangle=\mathbb{F}_p[x]/\langle\overline{f}(x)\rangle=\{\sum_{j=0}^{d-1}a_jx^j\mid a_0,a_1,\ldots,a_{d-1}\in\mathbb{F}_p\}$
and $\mathbb{F}_p[x]/\langle\overline{f}(x)\rangle={\mathcal T}$ as sets, one can
verify easily that ${\mathcal T}$ is a system of representatives for the equivalence classes of ${\mathcal R}$ under
congruence modulo $f(x)$. Then the conclusion follows from Lemma 2.3(i) immediately.

\par
  (iii) It follows from (ii), Lemma 2.2 and the general finite chain ring theory (see [14], for example).
\hfill $\Box$

\vskip 3mm \par
  Next, we consider how to decompose the ring $\mathbb{Z}_{p^s}[x]/\langle x^{N}-(1+pw)\rangle$
into a direct sum of finite chain rings, where $N=p^kn$ and ${\rm gcd}(p,n)=1$.

\vskip 3mm \noindent
  {\bf Lemma 2.6} \textit{Let $w\in \mathbb{Z}_{p^s}^{\times}$. Then ${\rm ord}(1+pw)=p^v$ for some positive integer $v$, and there exists a unique $w_0\in \mathbb{Z}_{p^s}^{\times}$
modulo $p^{s-1}$ such that $(1+pw_0)^n=1+pw$}.

\vskip 3mm \noindent
  \textit{ Proof} It is known that $1+p\mathbb{Z}_{p^s}$ is a multiplicative subgroup of $\mathbb{Z}_{p^s}^{\times}$
with order $p^{s-1}$ (cf. [16] Corollary 14.12]). Hence ${\rm ord}(1+pw)=p^v$ for some positive integer $v$. Since ${\rm gcd}(n,p)=1$, the mapping $1+pa\mapsto (1+pa)^n$ ($\forall a\in \mathbb{Z}_{p^s}$)
is an automorphism of the multiplicative group $1+p\mathbb{Z}_{p^s}$. Hence there is a unique
element $w_0\in \mathbb{Z}_{p^s}$ modulo $p^{s-1}$ such that $(1+pw_0)^n=1+pw$.

\par
   As every element of $\mathbb{Z}_{p^s}$ has a unique
$p$-expansion, there is a unique integer $k$, $0\leq k\leq s-1$, such that $w_0=p^kb$ for some $b\in \mathbb{Z}_{p^s}^{\times}$.
Suppose that $k\geq 1$, then $1+pw_0=1+p^{k+1}b$, which implies $(1+pw_0)^n=1+p^{k+1}bn+p^{2(k+1)}c=1+p(p^kbn+p^{2k+1}c)$ for some $c\in \mathbb{Z}_{p^s}$.
From this and by $(1+pw_0)^n=1+pw$ we deduce $pw=p(p^kbn+p^{2k+1}c)$, i.e., $w=p^kbn+p^{2k+1}c+p^{s-1}e$ for some $e\in \mathbb{Z}_{p^s}$,
and hence $w^s=0$, which contradicts that $w\in \mathbb{Z}_{p^s}^{\times}$. Therefore, we conclude that $k=0$ and hence
$w_0=b\in \mathbb{Z}_{p^s}^{\times}$.
\hfill $\Box$

\vskip 3mm \par
   In the rest of this paper, we adopt the following notations.

\vskip 3mm \noindent
  {\bf Notation 2.7} Let $w,w_0\in \mathbb{Z}_{p^s}^{\times}$ satisfying $(1+pw_0)^n=1+pw$, and assume
\begin{equation}
y^n-1=f_1(y)f_2(y)\ldots f_r(y),
\end{equation}
where $f_1(y),\ldots, f_r(y)$ are pairwise coprime monic basic irreducible polynomials
in $\mathbb{Z}_{p^s}[y]$. For each $i$, $1\leq i\leq r$, assume ${\rm deg}(f_i(y))=d_i$ and denote
$$\Gamma_i=\mathbb{Z}_{p^s}[y]/\langle f_i(y)\rangle \ {\rm and} \ {\mathcal R}_i=\mathbb{Z}_{p^s}[x]/\langle f_i(x^{p^k}(1+pw_0)^{-1})\rangle.$$

\vskip 3mm \par
  Let $1\leq i\leq r$. By Lemma 2.4(i) we know that $\Gamma_i$ is a Galois ring of characteristic $p^s$ and cardinality $p^{sd_i}$
and $\Gamma_i=\mathbb{Z}_{p^s}[\zeta_i]$, where $\zeta_i=y+\langle f_i(y)\rangle\in \Gamma_i$ satisfying
$\zeta_i^{p^{d_i}-1}=1$. By Theorem 2.5 we have the following corollary.

\vskip 3mm\noindent
   {\bf Corollary 2.8}  \textit{Denote}
${\mathcal T}_i=\{\sum_{j=0}^{d-1}a_jx^j\mid a_0,a_1,\ldots,a_{d_i-1}\in\mathbb{F}_p\}\subseteq {\mathcal R}_i$
\textit{in which we regard $\mathbb{F}_p$ as a subset of $\mathbb{Z}_{p^s}$. Then}

\vskip 2mm\par
   (i) \textit{${\mathcal R}_i$ is a finite chain ring with maximal ideal $\langle f_i(x)\rangle$
generated by $f_i(x)$, the nilpotency
index of $f_i(x)$ is equal to $p^ks$ and ${\mathcal R}_i/\langle f_i(x)\rangle$ is a finite field of cardinality $p^{d_i}$}.

\vskip 2mm\par
   (ii) \textit{Every element $\alpha$ of ${\mathcal R}_i$ has a unique $f_i(x)$-adic  expansion}:
\begin{center}
$\alpha=\sum_{j=0}^{p^ks-1}b_j(x)f_i(x)^j, \ b_0(x),b_1(x),\ldots,b_{p^ks-1}(x)\in {\mathcal T}_i.$
\end{center}

\vskip 2mm\par
   (iii) \textit{For each integer $l$, $1\leq l\leq p^ks$, let $\langle f_i(x)^l\rangle$ be the
ideal of ${\mathcal R}_i$ generated by $f_i(x)^l$. Then
$\langle f_i(x)^l\rangle=\{\sum_{j=l}^{p^ks-1}b_j(x)f_i(x)^j\mid b_l(x),\ldots,b_{p^ks-1}(x)\in {\mathcal T}_i\},$
and every element $\beta$ of the residue class ring ${\mathcal R}_i/\langle f_i(x)^l\rangle$ has a unique $f_i(x)$-adic expansion}:
$\beta=\sum_{j=0}^{l-1}b_j(x)f_i(x)^j, \ b_0(x),\ldots,b_{l-1}(x)\in {\mathcal T}_i.$

\par
\textit{Hence $\beta$ is an invertible element of ${\mathcal R}_i/\langle f_i(x)^l\rangle$,
i.e. $\beta\in ({\mathcal R}_i/\langle f_i(x)^l\rangle)^{\times}$,
if and only if $b_0(x)\neq 0$. Moreover,
$|f_i(x)^l{\mathcal R}_i|=|\langle f_i(x)^l\rangle|=p^{d_i(p^ks-l)}$ and $|({\mathcal R}_i/\langle f_i(x)^l\rangle)^{\times}|=(p^{d_i}-1)p^{(l-1)d_i}.$}

\vskip 3mm \par
  Let $1\leq i\leq r$ and denote $F_i(y)=\frac{y^n-1}{f_i(y)}\in \mathbb{Z}_{p^s}[y]$.
Since $F_i(y)$ and $f_i(y)$ are coprime, there are polynomials $a_i(y), b_i(y)\in \mathbb{Z}_{p^s}[y]$
such that
\begin{equation}
 a_i(y)F_i(y)+b_i(y)f_i(y)=1.
\end{equation}
Substituting $x^{p^k}(1+pw_0)^{-1}=\frac{x^{p^k}}{1+pw_0}$ for $y$ in (2.3) and (2.4), by $(1+pw_0)^n=1+pw$ we obtain
$$(1+pw_0)^{-n}(x^{p^kn}-(1+pw))
=(x^{p^k}(1+pw_0)^{-1})^n-1=\prod_{i=1}^rf_i(\frac{x^{p^k}}{1+pw_0})$$
and
$
a_i(\frac{x^{p^k}}{1+pw_0})F_i(\frac{x^{p^k}}{1+pw_0})+b_i(\frac{x^{p^k}}{1+pw_0})f_i(\frac{x^{p^k}}{1+pw_0})=1
$
in the ring $\mathbb{Z}_{p^s}[x]$, respectively. It is clear that
${\rm deg}(f_i(x^{p^k}(1+pw_0)^{-1}))=p^kd_i$ for $i=1,\ldots,r$.
In the rest of this paper, we set
\begin{equation}
\theta_i(x)\equiv a_i(\frac{x^{p^k}}{1+pw_0})F_i(\frac{x^{p^k}}{1+pw_0})=1-b_i(\frac{x^{p^k}}{1+pw_0})f_i(\frac{x^{p^k}}{1+pw_0})
\end{equation}
(mod $ x^{p^kn}-(1+pw)$). Then from Chinese Remainder Theorem for commutative rings with identity, we deduce that following theorem.

\vskip 3mm \noindent
   {\bf Theorem 2.9} \textit{Denote ${\mathcal A}=\mathbb{Z}_{p^s}[x]/\langle x^{p^kn}-(1+pw)\rangle$. We have the following}:

\vskip 2mm \par
   (i) \textit{$\theta_1(x)+\ldots+\theta_r(x)=1$, $\theta_i(x)^2=\theta_i(x)$
and $\theta_i(x)\theta_j(x)=0$ in ${\mathcal A}$ for all $1\leq i\neq j\leq r$}.

\vskip 2mm \par
   (ii) \textit{${\mathcal A}={\mathcal A}_1\oplus\ldots\oplus {\mathcal A}_r$, where ${\mathcal A}_i=\theta_i(x){\mathcal A}$ and its multiplicative
identity is $\theta_i(x)$. Moreover, this decomposition is a direct sum of rings
in that ${\mathcal A}_i{\mathcal A}_j=\{0\}$ for all $i$ and $j$, $1\leq i\neq j\leq r$}.

\vskip 2mm \par
   (iii) \textit{For each $1\leq i\leq r$, define a mapping $\tau_i: a(x)\mapsto \theta_i(x)a(x)$
$(\forall a(x)\in {\mathcal R}_i=\mathbb{Z}_{p^s}[x]/\langle f_i(x^{p^k}(1+pw_0)^{-1})\rangle)$. Then $\tau_i$ is a ring isomorphism
from ${\mathcal R}_i$ onto ${\mathcal A}_i$. Hence
$|{\mathcal A}_i|=p^{p^ksd_i}$}.

\vskip 2mm \par
   (iv) \textit{Define $\tau: (a_1(x),\ldots,a_r(x))\mapsto \tau_1(a_1(x))+\ldots+\tau_r(a_r(x))$, i.e.
$$\tau(a_1(x),\ldots,a_r(x))=\sum_{i=1}^r\theta_i(x)a_i(x) \ ({\rm mod} \ x^{p^kn}-(1+pw))$$
for all
$a_i(x)\in {\mathcal R}_i$, $i=1,\ldots,r$. Then $\tau$ is a ring isomorphism from ${\mathcal R}_1\times\ldots$ $\times{\mathcal R}_r$ onto
${\mathcal A}$}.

%%%%
%%%% Section 3
%%%%
\section{\bf Structure of $(1+pw)$-Constacyclic Codes over $\mathbb{Z}_{p^s}+u\mathbb{Z}_{p^s}$}
\par
    In this section, we list all distinct $(1+pw)$-constacyclic codes over the ring $\mathbb{Z}_{p^s}+u\mathbb{Z}_{p^s}$ ($u^2=0$) of length $N$,
i.e., all distinct ideals of the ring $(\mathbb{Z}_{p^s}+u\mathbb{Z}_{p^s})[x]/\langle x^{N}-(1+pw)\rangle$ where
$N=p^kn$ and $w\in \mathbb{Z}_{p^s}^{\times}$.

\par
  For any $\alpha\in (\mathbb{Z}_{p^s}+u\mathbb{Z}_{p^s})[x]/\langle x^{N}-(1+pw)\rangle$, $\alpha$ can be
expressed as
$$\alpha=\sum_{j=0}^{N-1}(a_j+b_ju)x^j, \ a_j,b_j\in \mathbb{Z}_{p^s}, \ j=0,1,\ldots,N-1$$
uniquely. Denote $\xi=\sum_{j=0}^{N-1}a_jx^j$ and $\eta=\sum_{j=0}^{N-1}b_jx^j$. Then $\xi,\eta\in {\mathcal A}
=\mathbb{Z}_{p^s}[x]/\langle x^{N}-(1+pw)\rangle$
and the map
$$\sigma: \alpha\mapsto \xi+\eta u, \ \forall \alpha\in  (\mathbb{Z}_{p^s}+u\mathbb{Z}_{p^s})[x]/\langle x^{N}-(1+pw)\rangle$$
is a ring isomorphism from $(\mathbb{Z}_{p^s}+u\mathbb{Z}_{p^s})[x]/\langle x^{N}-(1+pw)\rangle$ onto
${\mathcal A}+u{\mathcal A}$ ($u^2=0$).
In the rest of this paper, we will identify $(\mathbb{Z}_{p^s}+u\mathbb{Z}_{p^s})[x]/\langle x^{N}-(1+pw)\rangle$
with ${\mathcal A}+u{\mathcal A}$ under this isomorphism $\sigma$. Moreover, for any $1\leq i\leq r$ let
${\mathcal R}_i+u{\mathcal R}_i={\mathcal R}_i[u]/\langle u^2\rangle$,
where ${\mathcal R}_i=\mathbb{Z}_{p^s}[x]/\langle f_i(x^{p^k}(1+pw_0)^{-1})\rangle$.

\vskip 3mm \noindent
  {\bf Lemma 3.1} \textit{Let $1\leq i\leq r$. Using the notations of Theorem 2.9, for any
$a(x),b(x)\in {\mathcal R}_i$ we define}
\begin{eqnarray*}
\tau_i(a(x)+b(x)u) &=& \tau_i(a(x))+\tau_i(b(x))u \\
  &=&\theta_i(x)(a(x)+b(x)u) \ ({\rm mod} \ x^{p^kn}-(1+pw)).
\end{eqnarray*}
\textit{Then $\tau_i$ is a ring isomorphism from ${\mathcal R}_i+u{\mathcal R}_i$ onto ${\mathcal A}_i+u{\mathcal A}_i$ ($u^2=0$)}.

\vskip 3mm \noindent
   \textit{Proof} By Theorem 2.9(iii), the isomorphism $\tau_i:{\mathcal R}_i\rightarrow {\mathcal A}_i$ induces
an isomorphism of polynomial rings from ${\mathcal R}_i[u]$ onto ${\mathcal A}_i[u]$ in the natural
way that $\sum_{j}a_j(x)u^j\mapsto \sum_{j}\tau_i(a_j(x))u^j$ ($\forall a_j(x)\in {\mathcal R}_i$). Hence
$\tau_i$ is a ring isomorphism from ${\mathcal R}_i+u{\mathcal R}_i={\mathcal R}_i[u]/\langle u^2\rangle$ onto ${\mathcal A}_i+u{\mathcal A}_i
={\mathcal A}_i[u]/\langle u^2\rangle$.
\hfill $\Box$

\vskip 3mm \noindent
  {\bf Lemma 3.2} \textit{Using the notations above and the notations of Section 2,  The following statements are
equivalent}:

\vskip 2mm \par
   (i) \textit{${\mathcal C}$ is a $(1+pw)$-constacyclic code over $\mathbb{Z}_{p^s}+u\mathbb{Z}_{p^s}$ of length $p^kn$}.

\vskip 2mm \par
   (ii) \textit{${\mathcal C}$ is an ideal of ${\mathcal A}+u{\mathcal A}$}.

\vskip 2mm \par
   (iii) \textit{For each $1\leq i\leq r$, there is a unique ideal $C_i$ of
the ring ${\mathcal R}_i+u{\mathcal R}_i$ $(u^2=0)$ such that}
${\mathcal C}=\oplus_{i=1}^r\theta_i(x)C_i \ ({\rm mod} \ x^{p^kn}-(1+pw)).$

\vskip 3mm \noindent
 \textit{Proof} (i)$\Leftrightarrow$ (ii) It follows from the identification
of $(\mathbb{Z}_{p^s}+u\mathbb{Z}_{p^s})[x]/\langle x^{p^kn}-(1+pw)\rangle$
with ${\mathcal A}+u{\mathcal A}$.

\par
  (ii)$\Leftrightarrow$ (iii) By Theorem 2.9(ii) we have
${\mathcal A}=\oplus_{i=1}^r{\mathcal A}_i$. Hence
${\mathcal A}+u{\mathcal A}={\mathcal A}[u]/\langle u^2\rangle=\oplus_{i=1}^r({\mathcal A}_i[u]/\langle u^2\rangle)
=\oplus_{i=1}^r({\mathcal A}_i+u{\mathcal A}_i)$ and this decomposition is a direct sum of rings in that $({\mathcal A}_i+u{\mathcal A}_i)({\mathcal A}_j+u{\mathcal A}_j)=\{0\}$ for any $i,j$, $1\leq i\neq j\leq r$. Therefore, ${\mathcal C}$ is an ideal of ${\mathcal A}+u{\mathcal A}$
if and only if for each $1\leq i\leq r$, there is a unique ideal ${\mathcal C}_i$ of
the ring ${\mathcal A}_i+u{\mathcal A}_i$ such that ${\mathcal C}=\oplus_{i=1}^r{\mathcal C}_i$. From this and by Lemma 3.1, we deduce that
${\mathcal C}_i$ is an ideal of ${\mathcal A}_i+u{\mathcal A}_i$ if and only if there is a unique ideal $C_i$ of ${\mathcal R}_i+u{\mathcal R}_i$ such
that ${\mathcal C}_i=\tau_i(C_i)=\theta_i(x)C_i=\{\theta_i(x)c_i(x)\mid
c_i(x)\in C_i\}$ (mod $x^{p^kn}-(1+pw)$).
\hfill $\Box$

\vskip 3mm \par
  In order to present all $(1+pw)$-constacyclic codes over $\mathbb{Z}_{p^s}+u\mathbb{Z}_{p^s}$ of length $p^kn$, by Lemma 3.2
it is sufficient to give all ideals of
the ring ${\mathcal R}_i+u{\mathcal R}_i$ for all $i=1,\ldots,r$. Now, we give the following theorem.

\vskip 3mm \noindent
  {\bf Theorem 3.3} \textit{Using the notations above, in the rest of this paper for
any $1\leq l\leq p^{k}s$ we denote $({\mathcal R}_i[x]/\langle f_i(x)^l\rangle)^{\times}$ by $\Delta_l^{(i)}$, i.e.,}
$$\Delta_l^{(i)}=\{\sum_{j=0}^{l-1}b_j(x) f_i(x)^j\mid
b_j(x)\in {\mathcal T}_i, b_0(x)\neq 0, 0\leq j\leq l-1\}$$
\textit{Then all distinct
ideals $C_i$ of ${\mathcal R}_i+u{\mathcal R}_i$ are given by the following table}:
{\small\begin{center}
\begin{tabular}{llll}\hline
case &  number of ideals  &  $C_i$ (ideal of ${\mathcal R}_i+u{\mathcal R}_i$)   &   $|C_i|$  \\ \hline
I.   &  $p^{k}s+1$       & $\bullet$ $\langle f_i(x)^{l_i}\rangle$  $(0\leq l_i\leq p^{k}s)$ & $p^{2d_i(p^{k}s-l_i)}$ \\
II.   &   $p^{k}s$          & $\bullet$   $\langle uf_i(x)^{m_i}\rangle$   $(0\leq m_i\leq p^{k}s-1)$ &  $p^{d_i(p^{k}s-m_i)}$ \\
III. &  $\Omega(p^{d_i},p^{k}s)$ & $\bullet$  $\langle f_i(x)^{l_i}+uf_i(x)^{t_i}h(x)\rangle$ &  $p^{2d_i(p^{k}s-l_i)}$ \\
     &                   & $(h(x)\in \Delta_{l_i-t_i}^{(i)}$, $t_i\geq 2l_i-p^{k}s$, & \\
     &                   & $0\leq t_i<l_i\leq p^{k}s-1)$ \\
     &                  & $\bullet$  $\langle f_i(x)^{l_i}+uf_i(x)^{t_i}h(x)\rangle$ &  $p^{d_i(p^{k}s-t_i)}$ \\
     &                   & $(h(x)\in \Delta_{p^{k}s-l_i}^{(i)}$, $t_i< 2l_i-p^{k}s$, & \\
     &                   & $0\leq t_i<l_i\leq p^{k}s-1)$ & \\
IV.  &  $\frac{1}{2}p^{k}s(p^{k}s-1)$ &  $\bullet$ $\langle f_i(x)^{l_i},uf_i(x)^{m_i}\rangle$ &  $p^{d_i(2p^{k}s-(l_i+m_i))}$ \\
     &                     &  ($0\leq m_i<l_i\leq p^{k}s-1$)     & \\
V.  &  $(p^{d_i}-1)$ & $\bullet$  $\langle f_i(x)^{l_i}+uf_i(x)^{t_i}h(x),uf_i(x)^{m_i}\rangle$ &  $p^{d_i(2p^{k}s-(l_i+m_i))}$ \\
     & $\cdot \Psi(p^{d_i},p^{k}s)$                    & $(h(x)\in \Delta_{m_i-t_i}^{(i)}$,  &    \\
     &                     &   $l_i+m_i\leq p^{k}s+t_i-1$, & \\
     &                     &  $0\leq t_i<m_i<l_i\leq p^{k}s-1$)     &  \\ \hline
\end{tabular}
\end{center}}

\noindent
 \textit{where}
\begin{eqnarray*}
\Omega(p^{d_i},p^ks)&=&\frac{p^{d_i(\frac{p^ks}{2}+1)}+
p^{d_i\cdot\frac{p^ks}{2}}-2}{p^{d_i}-1}-(p^ks+1)\\
  &&+(p^{d_i}-1)\sum_{j=\frac{p^ks}{2}+1}^{p^ks-1}(2j-p^ks)p^{d_i(p^ks-j-1)}
\end{eqnarray*}
\textit{if $p^ks$ is even;}
\begin{eqnarray*}
\Omega(p^{d_i},p^ks)&=&\frac{2(p^{d_i\cdot\frac{p^ks+1}{2}}-1)}{p^{d_i}-1}-(p^ks+1)\\
  &&+(p^{d_i}-1)\sum_{j=\frac{p^ks+1}{2}}^{p^ks-1}(2j-p^ks)p^{d_i(p^ks-j-1)}
\end{eqnarray*}
\textit{if $p^ks$ is odd, and $\Psi(p^{d_i},p^{k}s)$ can be calculated by the following recurrence formula}:

\vskip 2mm\par
   \textit{$\Psi(p^{d_i},t)=0$ for $t=1,2,3$, $\Psi(p^{d_i},t)=1$ for $t=4$};

\vskip 2mm\par
  \textit{$\Psi(p^{d_i},t)=\Psi(p^{d_i},t-1)+\sum_{j=1}^{\lfloor\frac{t}{2}\rfloor-1}(t-2j-1)p^{d_i(j-1)}$ for $t\geq 5$}.

\vskip 2mm \noindent
  \textit{Therefore, the number of all
distinct ideals of the ring  ${\mathcal R}_i+u{\mathcal R}_i$ is equal to}
$$N_{(p,d_i,p^{k}s)}=\left\{\begin{array}{ll}\sum_{j=0}^\lambda(1+4j)p^{(\lambda-j)d_i} & {\rm if} \ p^{k}s=2\lambda; \cr & \cr
                                        \sum_{j=0}^\lambda(3+4j)p^{(\lambda-j)d_i} & {\rm if} \ p^{k}s=2\lambda+1. \end{array}\right.$$

\vskip 3mm \noindent
  \textit{Proof} We define a map $\varrho: {\mathcal R}_i+u{\mathcal R}_i\rightarrow {\mathcal R}_i$ by
$\varrho(\alpha+u\beta)=\alpha$ ($\forall \alpha,\beta\in {\mathcal R}_i$). Then $\varrho$ is a surjective ring
homomorphism from ${\mathcal R}_i+u{\mathcal R}_i$ onto ${\mathcal R}_i$.

\par
   Let $J$ be an ideal of ${\mathcal R}_i+u{\mathcal R}_i$ and $\varrho|_{J}$ be the restriction of $\varrho$ to $J$.
Then $\varrho|_{J}$ is a surjective ring homomorphism from $J$ onto $\varrho(J)=\{\varrho(\xi)\mid \xi\in J\}$,
which implies $\varrho(J)\cong J/{\rm Ker}(\varrho|_{J})$ where
${\rm Ker}(\varrho|_{J})=\{\xi\in J\mid \varrho(\xi)=0\}$ is the kernel of $\varrho|_{J}$. Therefore,
$|J|=|\varrho(J)||{\rm Ker}(\varrho|_{J})|$.

\par
  Let $(J:u)=\{\xi\in {\mathcal R}_i+u{\mathcal R}_i\mid u\xi\in J\}$. Then $(J:u)$ is an ideal of ${\mathcal R}_i+u{\mathcal R}_i$
satisfying $J\subseteq (J:u)$. Since $\varrho$ is a surjective ring
homomorphism, we see that $\varrho(J)$ and $\varrho(J:u)$ are ideals of ${\mathcal R}_i$
satisfying $\varrho(J)\subseteq\varrho(J:u)$. As ${\mathcal R}_i$ is a finite chain ring described in Section 2,
there is a unique pair $(l_i,m_i)$ of integers, $0\leq m_i\leq l_i\leq p^ks$, such that
\begin{equation}
\varrho(J)=f_i(x)^{l_i}{\mathcal R}_i \ {\rm and}
\ \varrho(J:u)=f_i(x)^{m_i}{\mathcal R}_i.
\end{equation}
By the definition of $\varrho$, we have
\begin{eqnarray*}
{\rm Ker}(\varrho|_{J})&=&\{u\beta\in J\mid \beta\in {\mathcal R}_i\}
=u\{\beta+u\gamma\mid u(\beta+u\gamma)\in J, \ \beta,\gamma\in {\mathcal R}_i\}\\
&=&u(J:u)=u\varrho(J:u),
\end{eqnarray*}
which implies $|{\rm Ker}(\varrho|_{J})|=|\varrho(J:u)|$. From this, by Equation (3.1) we deduce that $|J|=|f_i(x)^{l_i}{\mathcal R}_i||f_i(x)^{m_i}{\mathcal R}_i|$. Then by
Corollary 2.8(iii) we have
\begin{equation}
|J|=p^{d_i(p^ks-l_i)}p^{d_i(p^ks-m_i)}=p^{d_i(2p^ks-(l_i+m_i))}.
\end{equation}

\par
  If $m_i=p^ks$, then $l_i=p^ks$ and $J=\{0\}=\langle f_i(x)^{p^ks}\rangle$. In the following, we assume
$m_i\leq p^ks-1$. Then we have the following cases.

\par
  {\bf Case (i)} $m_i=l_i$ where $0\leq l_i\leq p^ks-1$.
  In this case, we have $|J|=p^{d_i(2p^ks-2l_i)}=p^{2d_i(p^ks-l_i)}$ by Equation (3.2).

\par
  By $f_i(x)^{l_i}\in \varrho(J)$ there exists $\alpha\in {\mathcal R}_i$ such that
$f_i(x)^{l_i}+u\alpha\in J$, which implies $\langle f_i(x)^{l_i}+u\alpha\rangle\subseteq J$.
Conversely, let $\xi\in J$. By $\varrho(J)=f_i(x)^{l_i}{\mathcal R}_i$, there exist
$\gamma,\beta\in {\mathcal R}_i$ such that $\xi=f_i(x)^{l_i}\gamma+u\beta$, which implies
$u(\beta-\gamma\alpha)=\xi-\gamma(f_i(x)^{l_i}+u\alpha)\in J$, and hence
$\beta-\gamma\alpha\in \varrho(J:u)$. But $\varrho(J:u)=f_i(x)^{l_i}{\mathcal R}_i$, there exists
$\delta\in {\mathcal R}_i$ such that $\beta-\gamma\alpha=f_i(x)^{l_i}\delta$.
Hence $\xi=\gamma(f_i(x)^{l_i}+u\alpha)+uf_i(x)^{l_i}\delta=(\gamma+u\delta)(f_i(x)^{l_i}+u\alpha)
\in \langle f_i(x)^{l_i}+u\alpha\rangle$. Therefore, $J=\langle f_i(x)^{l_i}+u\alpha\rangle$.

\par
  If $l_i=0$, then $f_i(x)^{l_i}+u\alpha=1+u\alpha$ which is an invertible element of ${\mathcal R}_i+u{\mathcal R}_i$
($u^2=0$). Hence $J={\mathcal R}_i+u{\mathcal R}_i=\langle f_i(x)^0\rangle$. In the following, we assume
that $1\leq l_i\leq p^ks-1$.

\par
   Since every element of ${\mathcal R}_i$ has a unique $f_i(x)$-expansion, by
$uf_i(x)^{l_i}=u(f_i(x)^{l_i}+u\alpha)\in J$ we may assume that
$\alpha=0$ or $\alpha=f_i(x)^{t_i}h(x)$, where $0\leq t_i< l_i$ and
$h(x)=\sum_{j=0}^{p^ks-t_i-1}h_j(x)f_i(x)^j$ with $h_0(x),\ldots,h_{p^ks-t_i-1}(x)\in {\mathcal T}_i$
satisfying $h_0(x)\neq 0$.

\par
  If $\alpha=0$, $J=\langle f_i(x)^{l_i}\rangle$ where $1\leq l_i\leq p^ks-1$.

\par
  Otherwise, we have $J=\langle f_i(x)^{l_i}+uf_i(x)^{t_i}h(x)\rangle$.
As $h_0(x)\neq 0$, $h(x)$ is an invertible element of ${\mathcal R}_i$ and
\begin{center}
$uf_i(x)^{p^ks-l_i+t_i}=f_i(x)^{p^ks-l_i}h(x)^{-1}(f_i(x)^{l_i}+uf_i(x)^{t_i}h(x))\in J$,
\end{center}
which implies $f_i(x)^{p^ks-l_i+t_i}\in \varrho(J:u)=f_i(x)^{l_i}{\mathcal R}_i$. Hence
$p^ks-l_i+t_i\geq l_i$, i.e., $t_i\geq 2l_i-p^ks$.

\par
   Now, let $h^{\prime}(x)=\sum_{j=0}^{p^ks-t_i^{\prime}-1}h_j^{\prime}(x)f_i(x)^j$ satisfy $J=\langle f_i(x)^{l_i}+uf_i(x)^{t_i^{\prime}}h^{\prime}(x)\rangle$, where $0\leq t_i^{\prime}<l_i$, $h_0^{\prime}(x),\ldots,h_{p^ks-t_i^{\prime}-1}^{\prime}(x)\in {\mathcal T}_i$
and $h_0^{\prime}(x)\neq 0$. Then
\begin{eqnarray*}
&&u(f_i^{t_i}(x)h(x)-f_i(x)^{t_i^{\prime}}h^{\prime}(x))\\
&=&(f_i(x)^{l_i}+uf_i(x)h(x))-(f_i(x)^{l_i}+uf_i(x)^{t_i^{\prime}}h^{\prime}(x))\in J,
\end{eqnarray*}
which implies $f_i^{t_i}(x)h(x)-f_i(x)^{t_i^{\prime}}h^{\prime}(x)\in f_i(x)^{l_i}{\mathcal R}_i$. As $t_i,t_i^{\prime}<l_i$, the
condition is equivalent to $t_i=t_i^{\prime}$ and $h(x)\equiv h^{\prime}(x)$ (mod $f_i(x)^{l_i-t_i}$).
Therefore, all distinct ideals of ${\mathcal R}_i+u{\mathcal R}_i$ are given by:
$J=\langle f_i(x)^{l_i}+uf_i(x)^{t_i}h(x)\rangle$, where $h(x)\in ({\mathcal R}_i/\langle f_i(x)^{l_i-t_i}\rangle)^{\times}
=\Delta_{l_i-t_i}^{(i)}$ and $t_i\geq 2l_i-p^ks$.

\par
  {\bf Case (ii)} $l_i=p^ks$ and $0\leq m_i\leq p^ks-1$.

\par
  In this case, we have $|J|=p^{d_i(2p^ks-(l_i+m_i))}=p^{d_i(p^ks-m_i)}$ by Equation (3.2) and $\varrho(J)=\{0\}$. By $\varrho(J)=\{0\}$
and $\varrho(J:u)=f_i(x)^{m_i}{\mathcal R}_i$, one can easily verify that $J=\langle uf_i(x)^{m_i}\rangle$.

\par
  {\bf Case (iii)} $m_i=0$ and $1\leq l_i\leq p^ks-1$.

\par
  In this case, we have $|J|=p^{d_i(2p^ks-(l_i+m_i))}=p^{d_i(2p^ks-l_i)}$ by Equation (3.2).
Moreover, by $1\in f_i(x)^{0}{\mathcal R}_i=\varrho(J:u)$ we conclude that $u\in J$. Then by $\varrho(J)=f_i(x)^{l_i}{\mathcal R}_i$ it follows that
$J=\langle f_i(x)^{l_i},u\rangle$ immediately.

\par
  {\bf Case (iv)} $1\leq m_i<l_i\leq p^ks-1$.

\par
   In this case, we have $|J|=p^{d_i(2p^ks-(l_i+m_i))}$ by Equation (3.2). By $\varrho(J)=f_i(x)^{l_i}{\mathcal R}_i$ and
$\varrho(J:u)=f_i(x)^{m_i}{\mathcal R}_i$ we have $uf_i(x)^{m_i}\in J$ and there exists $\alpha\in {\mathcal R}_i$ such
that $f_i(x)^{l_i}+u\alpha \in J$. It is obvious that $\langle f_i(x)^{l_i}+u\alpha, uf_i(x)^{m_i}\rangle\subseteq J$.
Conversely, let $\xi\in J$. By $\varrho(\xi)\in f_i(x)^{l_i}{\mathcal R}_i$ there exist
$\gamma,\beta\in {\mathcal R}_i$ such that $\xi=f_i(x)^{l_i}\gamma+u\beta$. Then by
$u(\beta-\gamma\alpha)=\xi-\gamma(f_i(x)^{l_i}+u\alpha)\in J$, it follows that
$\beta-\gamma\alpha\in \varrho(J:u)=f_i(x)^{m_i}{\mathcal R}_i$, which implies
$\beta-\gamma\alpha=f_i(x)^{m_i}\delta$ for some $\delta\in {\mathcal R}_i$. Hence
$\xi=\gamma(f_i(x)^{l_i}+u\alpha)+\delta f_i(x)^{m_i}\in \langle f_i(x)^{l_i}+u\alpha,uf_i(x)^{m_i}\rangle$.
Therefore, $J=\langle f_i(x)^{l_i}+u\alpha,uf_i(x)^{m_i}\rangle$.

\par
  If $\alpha=0$, we have $J=\langle f_i(x)^{l_i},uf_i(x)^{m_i}\rangle$ where $1\leq m_i<l_i\leq p^ks-1$.

\par
  Otherwise, an argument similar to the proof of Case (i) shows that:
$J=\langle f_i(x)^{l_i}+uf_i(x)^{t_i}h(x),uf_i(x)^{m_i}\rangle$ where
$h(x)\in ({\mathcal R}_i/\langle f_i(x)^{m_i-t_i}\rangle)^{\times}=\Delta_{m_i-t_i}^{(i)}$
and $0\leq t_i\leq m_i-1$.

\par
  By $uf_i(x)^{p^ks-l_i+t_i}=f_i(x)^{p^ks-l_i}h(x)^{-1}(f_i(x)^{l_i}+uf_i(x)^{t_i}h(x))\in J$, we see that
$f_i(x)^{p^ks-l_i+t_i}\in \varrho(J:u)=f_i(x)^{m_i}{\mathcal R}_i$, which implies  $p^ks-l_i+t_i\geq m_i$. Hence we have one of the following
two cases:

\par
  ($\diamondsuit$-1) $p^ks-l_i+t_i=m_i$, i.e., $m_i-t_i=p^ks-l_i$ or $l_i+m_i=p^ks+t_i$.

\par
  In this case, $\Delta_{m_i-t_i}^{(i)}=\Delta_{p^ks-l_i}^{(i)}$. Now, by
\begin{center}
  $uf_i(x)^{m_i}=f_i(x)^{p^ks-l_i}h(x)^{-1}(f_i(x)^{l_i}+uf_i(x)^{t_i}h(x))$
\end{center}
it follows that
$J=\langle f_i(x)^{l_i}+uf_i(x)^{t_i}h(x)\rangle$. Moreover, we have $2l_i>l_i+m_i=p^ks+t_i$, i.e., $t_i<2l_i-p^ks$. Therefore,
$h(x)\in \Delta_{p^ks-l_i}^{(i)}$ and $|J|=p^{d_i(2p^ks-(l_i+m_i))}=p^{d_i(2p^ks-(p^ks+t_i))}=p^{d_i(p^ks-t_i)}$.

\par
  ($\diamondsuit$-2) $p^ks-l_i+t_i>m_i$, i.e., $l_i+m_i\leq p^ks+t_i-1$.

\par
  In this case, $J=\langle f_i(x)^{l_i}+uf_i(x)^{t_i}h(x),uf_i(x)^{m_i}\rangle$ where
$h(x)\in \Delta_{m_i-t_i}^{(i)}$.

\vskip 2mm\par
  Therefore, all distinct ideals of ${\mathcal R}_i+u{\mathcal R}_i$ are given by
(I)--(V) and the number of elements in each ideal is given at the right side of the table.

\par
   It is obvious that the numbers of ideals in cases (I), (II) and (IV) are equal to $p^ks+1$, $p^ks$ and $\frac{1}{2}p^ks(p^ks-1)$ respectively. Now, we count the number of ideals in (III) and (V), respectively.

\par
   First, we consider ideals in (III).
Let $0\leq t_i<l_i\leq p^ks-1$. When $l_i\leq \lfloor \frac{p^ks+1}{2}\rfloor$, i.e., $2l_i\leq p^ks$, then
$t_i$ satisfies $t_i\geq 2l_i-p^ks$ for all $0\leq t_i\leq l_i-1$. In this case, the number of ideals is equal to
{\small
$$
N_1=\sum_{l_i=1}^{\lfloor \frac{p^ks+1}{2}\rfloor}\sum_{t_i=0}^{l_i-1}(p^{d_i}-1)p^{d_i(l_i-t_i-1)}
=\frac{p^{d_i(\lfloor \frac{p^ks+1}{2}\rfloor+1)}-1}{p^{d_i}-1}-\left(\left\lfloor \frac{p^ks+1}{2}\right\rfloor+1\right).
$$ }
When $l_i\geq \lfloor \frac{p^ks+1}{2}\rfloor+1$, i.e., $2l_i> p^ks$, then
$t_i$ satisfies $t_i\geq 2l_i-p^ks$ for all $2l_i-p^ks\leq t_i\leq l_i-1$, and $t_i$ satisfies $t_i<2l_i-p^ks$ for all $0\leq t_i\leq 2l_i-p^ks-1$.
In this case, the number of ideals is equal to
{\footnotesize \begin{eqnarray*}
N_2&=&\sum_{l_i=\lfloor \frac{p^ks+1}{2}\rfloor+1}^{p^ks-1}\left(\sum_{t_i=2l_i-p^ks}^{l_i-1}(p^{d_i}-1)p^{d_i(l_i-t_i-1)}+\sum_{t_i=0}^{2l_i-p^ks-1}(p^{d_i}-1)p^{d_i(p^ks-l_i-1)}\right)\\
&=&\frac{p^{d_i(p^ks-\lfloor \frac{p^ks+1}{2}\rfloor)}-1}{p^{d_i}-1}-\left(p^ks-\left\lfloor \frac{p^ks+1}{2}\right\rfloor\right)\\
  &&+\sum_{l_i=\lfloor \frac{p^ks+1}{2}\rfloor+1}^{p^ks-1}(2l_i-p^ks)(p^{d_i}-1)p^{d_i(p^ks-l_i-1)}.
\end{eqnarray*}}
Hence the number of ideals in (III) is equal to $\Omega(p^{d_i},p^ks)=N_1+N_2$.
    Next, we denote
$\Psi(p^{d_i},p^ks)=\sum_{t_i=0}^{p^ks-4}\sum_{m_i=t_i+1}^{\frac{p^ks+t_i}{2}-1}\sum_{l_i=m_i+1}^{p^ks+t_i-1-m_i}p^{d_i(m_i-t_i-1)}.$
   Then
the number of ideals in (V) is equal to $(p^{d_i}-1)\Psi(p^{d_i},p^ks)$, where
{\footnotesize
\begin{eqnarray*}
\Psi(p^{d_i},p^ks)&=&\sum_{t_i=1}^{p^ks-4}\sum_{m_i=t_i+1}^{\frac{p^ks+t_i}{2}-1}\sum_{l_i=m_i+1}^{p^ks+t_i-1-m_i}p^{d_i(m_i-t_i-1)}
   +\sum_{m_i=1}^{\frac{p^ks}{2}-1}\sum_{l_i=m_i+1}^{p^ks-1-m_i}p^{d_i(m_i-1)}\\
    &=&\sum_{t_i^{\prime}=0}^{(p^ks-1)-4}\sum_{m_i^{\prime}=t_i^{\prime}+1}^{\frac{(p^ks-1)+t_i^{\prime}}{2}-1}
    \sum_{l_i^{\prime}=m_i^{\prime}+1}^{(p^ks-1)+t_i^{\prime}-1-m_i^{\prime}}p^{d_i(m_i^{\prime}-t_i^{\prime}-1)}\\
   &&+\sum_{m_i=1}^{\lfloor \frac{p^ks}{2}\rfloor-1}(p^ks-2m_i-1)p^{d_i(m_i-1)}\\
   &=&\Psi(p^{d_i},p^ks-1)+\sum_{m_i=1}^{\lfloor \frac{p^ks}{2}\rfloor-1}(p^ks-2m_i-1)p^{d_i(m_i-1)}
\end{eqnarray*} }
for all $s\geq 5$.

\par
   If $1\leq p^ks\leq 3$, there is no triple $(t_i,m_i,l_i)$ of integers satisfying $0\leq t_i<m_i<l_i\leq p^ks-1$ and $l_i+m_i\leq p^ks+t_i-1$.
In this case, the number of ideals in (V) is equal to $0$. Then we set $\Psi(p^{d_i},p^ks)=0$ for all $p^ks\leq 3$.

\par
   If $p^ks=4$, there is a
unique triple $(t_i,m_i,l_i)=(0,1,2)$ of integers satisfying $0\leq t_i<m_i<l_i\leq p^ks-1$ and $l_i+m_i\leq p^ks+t_i-1$. In this case, all distinct ideals in (V) are
given by $\langle f_i(x)^2+uh(x),uf_i(x)\rangle$, where $h(x)\in \Delta_1^{(i)}={\mathcal T}_i\setminus\{0\}$ and $|{\mathcal T}_i\setminus\{0\}|=p^{d_i}-1$. Then
we set $\Psi(p^{d_i},4)=1$.

\par
  Therefore, the number $N_i$ of ideals of ${\mathcal R}_i+u{\mathcal R}_i$ is equal to
$$ p^ks+1+p^ks+\Omega(p^{d_i},p^ks)
+\frac{1}{2}p^ks(p^ks-1)+(p^{d_i}-1)\Psi(p^{d_i},p^ks),$$
i.e., $N_i=1+\frac{1}{2}p^ks(p^ks+3)+\Omega(p^{d_i},p^ks)+(p^{d_i}-1)\Psi(p^{d_i},p^ks)$.
\hfill $\Box$

\vskip 3mm\par
   Then from Lemma 3.2 and Theorem 3.3 we deduce the following corollary.

\vskip 3mm\noindent
   {\bf Corollary 3.4} \textit{Every $(1+pw)$-constacyclic
code ${\mathcal C}$ over $\mathbb{Z}_{p^s}+u\mathbb{Z}_{p^s}$ of length $p^kn$ can be constructed by the following two steps}:

\vskip2 mm\par
  (i) \textit{For each $i=1,\ldots,r$, choose an ideal $C_i$ of ${\mathcal R}_i+u{\mathcal R}_i$
listed in Theorem 3.3}.

\vskip2 mm\par
  (ii) \textit{Set ${\mathcal C}=\oplus_{i=1}^r\theta_i(x)C_i=\sum_{i=1}^r\theta_i(x)C_i$ $({\rm mod} \ x^{p^kn}-(1+pw))$. Then the codewords in ${\mathcal C}$ is
equal to $|{\mathcal C}|=\prod_{i=1}^r|C_i|$}.

\vskip2 mm\noindent
 \textit{Using the notations of Theorem 3.3, the number of $(1+pw)$-constacyclic
codes over $\mathbb{Z}_{p^s}+u\mathbb{Z}_{p^s}$ of length $p^kn$ is equal to $\prod_{i=1}^rN_i$}.

\vskip 3mm\par
   Using the notations of Corollary 3.4, ${\mathcal C}=\oplus_{i=1}^r\theta_i(x)C_i$ is called the \textit{canonical form decomposition} of the $(1+pw)$-constacyclic
code ${\mathcal C}$ over $\mathbb{Z}_{p^s}+u\mathbb{Z}_{p^s}$.

\section{\bf Dual Codes of $(1+pw)$-Constacyclic Codes}
  In this section, we give the dual code of every $(1+pw)$-constacyclic code over $\mathbb{Z}_{p^s}+u\mathbb{Z}_{p^s}$
of length $N$ and investigate the self-duality of these codes.

\vskip 3mm
\noindent
  {\bf Lemma 4.1} \textit{Let $w\in \mathbb{Z}_{p^s}^{\times}$ and denote $\widehat{w}=-w(1+pw)^{-1}$.
Then $\widehat{w}\in \mathbb{Z}_{p^s}^{\times}$ and $(1+pw)^{-1}=1+p\widehat{w}$}.

\vskip 3mm
\noindent
  \textit{Proof}  Obviously, we have $\widehat{w}=-w(1+pw)^{-1}\in \mathbb{Z}_{p^s}^{\times}$ and $(1+pw)^{-1}=(1+pw-pw)(1+pw)^{-1}=1+p(w(1+pw)^{-1})$.
\hfill $\Box$

\vskip 3mm \par
  Using the notations of Lemma 4.1, by Lemma 1.2 we know that the dual code of each $(1+pw)$-constacyclic code
over $\mathbb{Z}_{p^s}+u\mathbb{Z}_{p^s}$ of length $p^kn$ is a $(1+p\widehat{w})$-constacyclic code
over $\mathbb{Z}_{p^s}+u\mathbb{Z}_{p^s}$ of length $p^kn$, i.e., an ideal of the ring
$(\mathbb{Z}_{p^s}+u\mathbb{Z}_{p^s})[x]/\langle x^{p^kn}-(1+p\widehat{w})\rangle\rangle$. In the following, we will determining
the dual code of each $(1+p\widehat{w})$-constacyclic code
over $\mathbb{Z}_{p^s}+u\mathbb{Z}_{p^s}$ of length $p^kn$ from its canonical form decomposition
given by Theorem 3.3 and Corollary 3.4.

\par
   By Lemmas 2.6 and 4.1, we can select a fixed
$\widehat{w}_0\in \mathbb{Z}_{p^s}^{\times}$ satisfying
$$(1+p\widehat{w}_0)^n=1+p\widehat{w} \ {\rm and} \ (1+pw_0)^{-1}=1+p\widehat{w}_0$$
in the rest of the paper. First, from Theorem 2.5 we deduce

\vskip 3mm\noindent
  {\bf Corollary 4.2} \textit{Using the notations of Notation 2.7, we denote
$$\widehat{{\mathcal R}}_i=\mathbb{Z}_{p^s}[x]/\langle f_i(x^{p^k}(1+p\widehat{w}_0)^{-1})\rangle$$
and ${\mathcal T}_i=\{\sum_{j=0}^{d_i-1}a_jx^j\mid a_0,a_1,\ldots,a_{d_i-1}\in \mathbb{F}_p\}\subseteq \widehat{{\mathcal R}}_i$.}

\vskip 2mm\par
   (i) \textit{$\widehat{{\mathcal R}}_i$ is a finite chain ring with maximal ideal $\langle f_i(x)\rangle$
generated by $f_i(x)$, the nilpotency
index of $f_i(x)$ is equal to $p^ks$ and $\widehat{{\mathcal R}}_i/\langle f_i(x)\rangle$ is a finite field of cardinality $p^{d_i}$}.

\vskip 2mm\par
   (ii) \textit{Every element $\alpha$ of $\widehat{{\mathcal R}}_i$ has a unique $f_i(x)$-adic  expansion}:
\begin{center}
$\alpha=\sum_{j=0}^{p^ks-1}b_j(x)f_i(x)^j, \ b_0(x),b_1(x),\ldots,b_{p^ks-1}(x)\in {\mathcal T}_i.$
\end{center}

\vskip 3mm\par
  Substituting $\frac{x^{p^k}}{1+p\widehat{w}_0}$ for $y$ in (2.3) and (2.4), we obtain
$$\frac{x^{p^kn}-(1+p\widehat{w})}{(1+p\widehat{w}_0)^n}=\left(\frac{x^{p^k}}{1+p\widehat{w}_0}\right)^n-1
=\prod_{i=1}^rf_i(\frac{x^{p^k}}{1+p\widehat{w}_0})$$
and
$
a_i(\frac{x^{p^k}}{1+p\widehat{w}_0})F_i(\frac{x^{p^k}}{1+p\widehat{w}_0})+b_i(\frac{x^{p^k}}{1+p\widehat{w}_0})f_i(\frac{x^{p^k}}{1+p\widehat{w}_0})=1
$
in the ring $\mathbb{Z}_{p^s}[x]$, respectively. It is clear that
${\rm deg}(f_i(x^{p^k}(1+p\widehat{w}_0)^{-1}))=p^kd_i$ for $i=1,\ldots,r$.
In the rest of this paper, we set
\begin{equation}
\widehat{\theta}_i(x)\equiv a_i(\frac{x^{p^k}}{1+p\widehat{w}_0})F_i(\frac{x^{p^k}}{1+p\widehat{w}_0})=1-b_i(\frac{x^{p^k}}{1+p\widehat{w}_0})f_i(\frac{x^{p^k}}{1+p\widehat{w}_0})
\end{equation}
(mod $x^{p^kn}-(1+p\widehat{w})$). Then by Chinese remainder theorem for commutative rings with identity, paralleling to Theorem 2.9 we have the following

\vskip 3mm \noindent
   {\bf Corollary 4.3} \textit{Denote $\widehat{{\mathcal A}}=\mathbb{Z}_{p^s}[x]/\langle x^{p^kn}-(1+p\widehat{w})\rangle$. We have the following}:

\vskip 2mm \par
   (i) \textit{$\widehat{\theta}_1(x)+\ldots+\widehat{\theta}_r(x)=1$, $\widehat{\theta}_i(x)^2=\widehat{\theta}_i(x)$
and $\widehat{\theta}_i(x)\widehat{\theta}_j(x)=0$ in $\widehat{{\mathcal A}}$ for all $1\leq i\neq j\leq r$}.

\vskip 2mm \par
   (ii) \textit{$\widehat{{\mathcal A}}=\widehat{{\mathcal A}}_1\oplus\ldots\oplus \widehat{{\mathcal A}}_r$, where $\widehat{{\mathcal A}}_i=\widehat{\theta}_i(x)\widehat{{\mathcal A}}$ and its multiplicative
identity is $\widehat{\theta}_i(x)$. Moreover, this decomposition is a direct sum of rings
in that $\widehat{{\mathcal A}}_i\widehat{{\mathcal A}}_j=\{0\}$ for all $i$ and $j$, $1\leq i\neq j\leq r$}.

\vskip 2mm \par
   (iii) \textit{For each $1\leq i\leq r$, define a mapping $\widehat{\tau}_i: a(x)\mapsto \widehat{\theta}_i(x)a(x)$
$(\forall a(x)\in \widehat{{\mathcal R}}_i=\mathbb{Z}_{p^s}[x]/\langle f_i(x^{p^k}(1+p\widehat{w}_0)^{-1})\rangle)$. Then $\widehat{\tau}_i$ is a ring isomorphism
from $\widehat{{\mathcal R}}_i$ onto $\widehat{{\mathcal A}}_i$ and can be extended to a ring isomorphism
from $\widehat{{\mathcal R}}_i+u \widehat{{\mathcal R}}_i$ onto $\widehat{{\mathcal A}}_i+u\widehat{{\mathcal A}}_i$ in the natural way}.

\vskip 2mm \par
   (iv) \textit{Define $\widehat{\tau}: (a_1(x),\ldots,a_r(x))\mapsto \widehat{\tau}_1(a_1(x))+\ldots+\widehat{\tau}_r(a_r(x))$, i.e.
$$\widehat{\tau}(a_1(x),\ldots,a_r(x))=\sum_{i=1}^r\widehat{\theta}_i(x)a_i(x) \ ({\rm mod} \ x^{p^kn}-(1+p\widehat{w}))$$
for all
$a_i(x)\in \widehat{{\mathcal R}}_i$, $i=1,\ldots,r$. Then $\widehat{\tau}$ is a ring isomorphism from $\widehat{{\mathcal R}}_1\times\ldots\times\widehat{{\mathcal R}}_r$ onto
$\widehat{{\mathcal A}}$}.

\vskip 3mm \par
  As in Section 3, we can identify $(\mathbb{Z}_{p^s}+u\mathbb{Z}_{p^s})[x]/\langle x^N-(1+p\widehat{w})\rangle$
with $\widehat{{\mathcal A}}+u\widehat{{\mathcal A}}$ ($u^2=0$). Paralleling to Sections 3, we can obtain conclusions for
$(1+p\widehat{w})$-constacyclic codes over $\mathbb{Z}_{p^s}+u\mathbb{Z}_{p^s}$
of length $N$, i.e., ideals of the ring $\widehat{{\mathcal A}}+u\widehat{{\mathcal A}}$. We omit these conclusions here for space saving.

\par
   Now, let $\alpha=(\alpha_0,\alpha_1,\ldots,\alpha_{N-1}), \beta=(\beta_0,\beta_1,\ldots,\beta_{N-1})\in (\mathbb{Z}_{p^s}+u\mathbb{Z}_{p^s})^N$,
where $N=p^ks$ and $\alpha_j,\beta_j\in \mathbb{Z}_{p^s}+u\mathbb{Z}_{p^s}$ for all $j=0,1\ldots,N-1$. In the rest of this paper,
we denote
$$\alpha(x)=\sum_{j=0}^{N-1}\alpha_jx^j\in (\mathbb{Z}_{p^s}+u\mathbb{Z}_{p^s})[x]/\langle x^N-(1+pw)\rangle={\mathcal A}+u{\mathcal A},$$
$$\beta(x)=\sum_{j=0}^{N-1}\beta_jx^j\in (\mathbb{Z}_{p^s}+u\mathbb{Z}_{p^s})[x]/\langle x^N-(1+p\widehat{w})\rangle=\widehat{{\mathcal A}}+u\widehat{{\mathcal A}}.$$
Recall that
the usual \textit{Euclidian inner product} of $\alpha$ and $\beta$ is defined by
$[\alpha,\beta]_E=\sum_{j=0}^{N-1}\alpha_j\beta_j\in \mathbb{Z}_{p^s}+u\mathbb{Z}_{p^s}$.
Let ${\mathcal C}$ be a linear code over $\mathbb{Z}_{p^s}+u\mathbb{Z}_{p^s}$ of length $N$, i.e.,
a $(\mathbb{Z}_{p^s}+u\mathbb{Z}_{p^s})$-submodule of $(\mathbb{Z}_{p^s}+u\mathbb{Z}_{p^s})^N$. The \textit{Euclidian dual code}
of ${\mathcal C}$ is defined by ${\mathcal C}^{\bot_E}=\{\alpha\in (\mathbb{Z}_{p^s}+u\mathbb{Z}_{p^s})^N\mid [\alpha,\beta]_E=0, \ \forall
\beta\in {\mathcal C}\}$, and ${\mathcal C}$ is said to be \textit{self-dual} if ${\mathcal C}={\mathcal C}^{\bot_E}$.

\par
   From now on, we define a mapping $\mu: {\mathcal A}\rightarrow\widehat{{\mathcal A}}$ by
$$\mu(a(x))=a(x^{-1})=\sum_{j=0}^{N-1}a_jx^{-j}\in \widehat{{\mathcal A}}, \ \forall a(x)=\sum_{j=0}^{N-1}a_jx^j\in {\mathcal A}.$$
Then one can easily verify that $\mu$ is a ring isomorphism from ${\mathcal A}$ onto $\widehat{{\mathcal A}}$.
Precisely, by $x^N=(1+p\widehat{w})$ in $\widehat{{\mathcal A}}$ and Lemma 4.1 it follows that
$$a(x^{-1})=a_0+\frac{x^N}{1+p\widehat{w}}\sum_{j=1}^{N-1}a_jx^{-j}=a_0+(1+pw)\sum_{j=1}^{N-1}a_jx^{N-j},$$
and the inverse $\mu^{-1}: \widehat{{\mathcal A}}\rightarrow {\mathcal A}$ of $\mu$ is given by
$$\mu^{-1}(b(x))=b(x^{-1})=b_0+\frac{x^N}{1+pw}\sum_{j=1}^{N-1}b_jx^{-j}=b_0+(1+p\widehat{w})\sum_{j=1}^{N-1}b_jx^{N-j}\in {\mathcal A},$$
for all $b(x)=\sum_{j=0}^{N-1}b_jx^j\in \widehat{{\mathcal A}}$ where $b_j\in \mathbb{Z}_{p^s}$.

\par
   For notations simplicity, we still denote $\mu^{-1}$ by
$\mu$. Now, $\mu$ can be extended to a ring isomorphism between ${\mathcal A}+u{\mathcal A}$ and $\widehat{{\mathcal A}}+u\widehat{{\mathcal A}}$
 in the natural way that
$$\mu: \xi_1+u\xi_2\mapsto \mu(\xi_1)+u\mu(\xi_2), \ \forall \xi_1,\xi_2\in {\mathcal A}$$
and
$\mu: \eta_1+u\eta_2\mapsto \mu(\eta_1)+u\mu(\eta_2), \ \forall \eta_1,\eta_2\in \widehat{{\mathcal A}}$,
respectively.

\par
  Using the notations above, by a direct calculation we get the following

\vskip 3mm \noindent
  {\bf Lemma 4.4} \textit{Let $\alpha,\beta\in (\mathbb{Z}_{p^s}+u\mathbb{Z}_{p^s})^N$.
Then $[\alpha,\beta]_E=0$  if $\alpha(x)\mu(\beta(x))=0$
in $(\mathbb{Z}_{p^s}+u\mathbb{Z}_{p^s})[x]/\langle x^N-(1+pw)\rangle$, where $\alpha(x)\in {\mathcal A}+u{\mathcal A}$
and $\beta(x)\in \widehat{{\mathcal A}}+u\widehat{{\mathcal A}}$}.

\vskip 3mm \par
   For any polynomial $f(y)=\sum_{j=0}^dc_jy^j\in \mathbb{Z}_{p^s}[y]$ of degree $d\geq 1$, recall that
the \textit{reciprocal polynomial} of $f(y)$ is defined as $\widetilde{f}(y)=y^df(\frac{1}{y})=\sum_{j=0}^dc_jy^{d-j}$, and
 $f(y)$ is said to be \textit{self-reciprocal} if $\widetilde{f}(y)=\delta f(y)$ for some $\delta\in \mathbb{Z}_{p^s}^{\times}$. Then by Equation (2.3) in Section 2, we have
$$y^n-1=-\widetilde{f}_1(y)\widetilde{f}_2(y)\ldots \widetilde{f}_r(y).$$
Since $f_1(y),f_2(y),\ldots,f_r(y)$ are pairwise coprime monic basic irreducible polynomials in $\mathbb{Z}_{p^s}[y]$,
it is known that $\widetilde{f}_1(y),\widetilde{f}_2(y),\ldots, \widetilde{f}_r(y)$  are pairwise coprime monic basic polynomials in $\mathbb{Z}_{p^s}[y]$ as well. Hence for each integer $i$, $1\leq i\leq r$,
there is a unique integer $i^{\prime}$, $1\leq i^{\prime}\leq r$, such that
$$\widetilde{f}_i(y)=\delta_if_{i^{\prime}}(y) \
{\rm where} \ \delta_i\in \mathbb{Z}_{p^s}^{\times}.$$
Since
$x^{p^kn}=1+p\widehat{w}=(1+p\widehat{w}_0)^n$ in $\widehat{{\mathcal A}}$, we see that
$$z^n=1 \ {\rm in} \ \widehat{{\mathcal A}}, \ {\rm where} \ z=x^{p^k}(1+p\widehat{w}_0)^{-1}\in \widehat{{\mathcal A}}.$$
Then by Equation (2.5) in Section 2 and $(1+pw_0)^{-1}=1+p\widehat{w}_0$, we have
\begin{eqnarray*}
 \mu(\theta_i(x))&=&1-b_i(x^{-p^k}(1+pw_0)^{-1})f_i(x^{-p^k}(1+pw_0)^{-1})\\
   &=&1-b_i((x^{p^k}(1+p\widehat{w}_0)^{-1})^{-1})f_i((x^{p^k}(1+p\widehat{w}_0)^{-1})^{-1})\\
   &=&1-b_i(z^{-1})f_i(z^{-1})\\
   &=&1-z^{n-{\rm deg}(b_i(y))-d_i}(z^{{\rm deg}(b_i(y))}b_i(z^{-1}))(z^{d_i}f_i(z^{-1}))\\
   &=&1-z^{n-{\rm deg}(b_i(y))-d_i}\widetilde{b}_i(z)\widetilde{f}_i(z)\\
   &=&1-\delta_iz^{n-{\rm deg}(b_i(y))-d_i}\widetilde{b}_i(z)f_{i^{\prime}}(z)\\
   &=&1-g_i(x^{p^k}(1+p\widehat{w}_0)^{-1})f_{i^{\prime}}(x^{p^k}(1+p\widehat{w}_0)^{-1})
\end{eqnarray*}
where $g_i(z)=\delta_iz^{n-{\rm deg}(b_i(y))-d_i}\widetilde{b}_i(z)\in \mathbb{Z}_{p^s}[z]$.
Similarly, we can prove that
$\mu(\theta_i(x))=h_i(x^{-p^k}(1+p\widehat{w}_0)^{-1})F_i(x^{-p^k}(1+p\widehat{w}_0)^{-1})$ for some
$h_i(z)\in \mathbb{Z}_{p^s}[z]$.
From these and by Equation (11), we deduce that
$\mu(\theta_i(x))=\widehat{\theta}_{i^{\prime}}(x).$

\par
   Therefore,
for each $1\leq i\leq r$ there is a unique integer $i^{\prime}$, $1\leq i^{\prime}\leq r$, such that $\mu(\theta_i(x))=
\widehat{\theta}_{i^{\prime}}(x)$. We still use $\mu$ to denote this map $i\mapsto i^{\prime}$; i.e.,
$$\mu(i)=i^{\prime} \ {\rm and} \ \mu(\theta_i(x))=\widehat{\theta}_{\mu(i)}(x).$$
Whether $\mu$ denotes the ring isomorphism between ${\mathcal A}+u{\mathcal A}$ and $\widehat{{\mathcal A}}+u\widehat{{\mathcal A}}$  or this map on the set $\{1,\ldots,r\}$ is determined by context.
The next lemma shows the compatibility of the two uses of $\mu$.

\vskip 3mm \noindent
  {\bf Lemma 4.5} \textit{With the notations above, we have the following}:

\vskip 2mm \par
   (i) \textit{$\mu$ is a permutation on $\{1,\ldots,r\}$ satisfying $\mu^{-1}=\mu$}.

\vskip 2mm \par
   (ii) \textit{After a rearrangement of $\theta_1(x),\ldots,\theta_r(x)$ there are integers $\lambda,\epsilon$ such that
$\mu(i)=i$ for all $i=1,\ldots,\lambda$ and $\mu(\lambda+j)=\lambda+\epsilon+j$ for all $j=1,\ldots,\epsilon$, where $\lambda\geq 1,
\epsilon\geq 0$ and $\lambda+2\epsilon=r$}.

\vskip 2mm\par
   (iii) \textit{For each integer $i$, $1\leq i\leq r$, there is a unique invertible element $\delta_i$ of
$\mathbb{Z}_{p^s}$ such that $\widetilde{f}_i(y)=\delta_i f_{\mu(i)}(y)$}.

\vskip 2mm \par
   (iv) \textit{For any integer $i$, $1\leq i\leq r$, $\mu(\theta_i(x))=\widehat{\theta}_{\mu(i)}(x)$,
$\mu(\widehat{\theta}_i(x))=\theta_{\mu(i)}(x)$,
$\mu({\mathcal A}_{i})=\widehat{{\mathcal A}}_{\mu(i)}$ and $\mu(\widehat{{\mathcal A}}_{i})={\mathcal A}_{\mu(i)}$. Then $\mu$ induces a ring isomorphism between ${\mathcal A}_{i}+u{\mathcal A}_{i}$
and $\widehat{{\mathcal A}}_{\mu(i)}+u\widehat{{\mathcal A}}_{\mu(i)}$}.

\vskip 3mm \noindent
  \textit{Proof} (i)--(iii) follow from the definition of the map $\mu$.

\par
   (iv) By the definition of $\mu$, we have $\mu(\theta_i(x))=\widehat{\theta}_{\mu(i)}(x)$ and
$\mu(\widehat{\theta}_i(x))=\theta_{\mu(i)}(x)$. Then the other conclusions follow from that ${\mathcal A}_i=\theta_i(x){\mathcal A}$, $\widehat{{\mathcal A}}_{\mu(i)}=\widehat{\theta}_{\mu(i)}(x)\widehat{{\mathcal A}}$,
$\widehat{{\mathcal A}}_i=\widehat{\theta}_i(x)\widehat{{\mathcal A}}$, ${\mathcal A}_{\mu(i)}=\theta_{\mu(i)}(x){\mathcal A}$,
${\mathcal A}_{i}+u{\mathcal A}_{i}={\mathcal A}_{i}[u]/\langle u^2\rangle$ and $\widehat{{\mathcal A}}_{\mu(i)}+u\widehat{{\mathcal A}}_{\mu(i)}
=\widehat{{\mathcal A}}_{\mu(i)}[u]/\langle u^2\rangle$, immediately.
\hfill $\Box$

\vskip 3mm \noindent
  {\bf Lemma 4.6} \textit{Let $1\leq i\leq r$. Denote by $\mu|_{{\mathcal A}_i}$ the restriction
of $\mu$ to ${\mathcal A}_i$ and define $\mu_i: {\mathcal R}_i\rightarrow \widehat{{\mathcal R}}_{\mu(i)}$ by
$\mu_i(c(x))=c(x^{-1})=\sum_{j=0}^{p^kd_i-1}c_jx^{-j}$ (for all $c(x)=\sum_{j=0}^{p^kd_i-1}c_jx^j$ with $c_j\in \mathbb{Z}_{p^s}$)}.
\textit{Then $\mu_i$ is a ring isomorphism from ${\mathcal R}_i$ onto $\widehat{{\mathcal R}}_{\mu(i)}$ such that the following
diagram commutes}
{\small $$\begin{array}{ccc}{\mathcal R}_i=\mathbb{Z}_{p^s}[x]/\langle f_i(x^{p^k}(1+pw_0)^{-1})\rangle & \stackrel{\mu_i}{\longrightarrow} &  \widehat{{\mathcal R}}_{\mu(i)}=\mathbb{Z}_{p^s}[x]/\langle f_{\mu(i)}(x^{p^k}(1+p\widehat{w}_0)^{-1})\rangle \cr
  \tau_i  \downarrow &  & \ \ \ \downarrow \widehat{\tau}_{\mu(i)} \cr
 \ \ \ \ {\mathcal A}_i & \stackrel{\mu|{{\mathcal A}_i}}{\longrightarrow} & \widehat{ {\mathcal A}}_{\mu(i)}
\end{array}$$}
\textit{Moreover, $\mu_i$ can be extended to a ring isomorphism from ${\mathcal R}_i+u{\mathcal R}_i$
onto $\widehat{{\mathcal R}}_{\mu(i)}+u\widehat{{\mathcal R}}_{\mu(i)}$ by the natural way:
$\mu_i: \alpha+u\beta\mapsto \mu_i(\alpha)+u\mu_i(\beta)$ for all $\alpha,\beta\in {\mathcal R}_i$}.

\vskip 3mm \noindent
  \textit{Proof} For any $c(x)\in {\mathcal R}_i$, by Theorem 2.9(iii), Corollary 4.3(iii) and $\mu(\theta_i(x))$ $=\widehat{\theta}_{\mu(i)}(x)=1-g_i(x^{p^k}(1+p\widehat{w}_0)^{-1})f_{\mu(i)}(x^{p^k}(1+p\widehat{w}_0)^{-1}),$
it follows that
\begin{eqnarray*}
&&\left((\widehat{\tau}_{\mu(i)})^{-1}\mu|_{{\mathcal A}_i}\tau_i\right)(c(x))\\
&=&\left((\widehat{\tau}_{\mu(i)})^{-1}\mu|_{{\mathcal A}_i}\right)\left(\theta_i(x)c(x)\right)=(\widehat{\tau}_{\mu(i)})^{-1}\left(\mu(\theta_i(x))c(x^{-1})\right) \\
&=&(\widehat{\tau}_{\mu(i)})^{-1}\left((1-g_i(x^{p^k}(1+p\widehat{w}_0)^{-1})f_{\mu(i)}(x^{p^k}(1+p\widehat{w}_0)^{-1}))c(x^{-1})\right)\\
&=&\left(1-g_i(x^{p^k}(1+p\widehat{w}_0)^{-1})f_{\mu(i)}(x^{p^k}(1+p\widehat{w}_0)^{-1})\right)c(x^{-1})  \\
&=& c(x^{-1}) \ \left({\rm mod} \ f_{\mu(i)}(x^{p^k}(1+p\widehat{w}_0)^{-1})\right),
\end{eqnarray*}
which implies $\mu_i(c(x))=((\widehat{\tau}_{\mu(i)})^{-1}\mu|_{{\mathcal A}_i}\tau_i)(c(x))$ for all $c(x)\in {\mathcal R}_i$. Hence
$\mu_i=(\widehat{\tau}_{\mu(i)})^{-1}\mu|_{{\mathcal A}_i}\tau_i$, which is a ring isomorphism from
${\mathcal R}_i$ onto $\widehat{{\mathcal R}}_{\mu(i)}$ such that the diagram commutes.

\par
  Obviously, $\mu_i$ can be extended to a ring isomorphism from ${\mathcal R}_i[u]$ onto $\widehat{{\mathcal R}}_{\mu(i)}[u]$ in
the natural way that $\mu_i:\sum_j\alpha_ju^j\mapsto \sum_j\mu_i(\alpha_j)u^j$ ($\forall \alpha_j\in {\mathcal R}_i$). Therefore,
$\alpha+u\beta\mapsto \mu_i(\alpha)+u\mu_i(\beta)$ $(\forall \alpha,\beta\in{\mathcal R}_i)$ is a
ring isomorphism from ${\mathcal R}_{i}[u]/\langle u^2\rangle={\mathcal R}_{i}+u{\mathcal R}_{i}$
 onto $\widehat{{\mathcal R}}_{\mu(i)}[u]/\langle u^2\rangle=\widehat{{\mathcal R}}_{\mu(i)}+u\widehat{{\mathcal R}}_{\mu(i)}$.
\hfill $\Box$

\vskip 3mm \noindent
  {\bf Lemma 4.7} \textit{Using the notations of Theorem 2.9{\rm (iv)} and Corollary 4.3{\rm (iv)}, let $\alpha(x)=\sum_{i=1}^r\theta_i(x)a_i(x)\in {\mathcal A}+u{\mathcal A}$ and $\beta(x)=\sum_{i=1}^r\widehat{\theta}_i(x)b_i(x)\in \widehat{{\mathcal A}}+u\widehat{{\mathcal A}}$,
where $a_i(x)\in{\mathcal R}_i+u{\mathcal R}_i$ and $b_i(x)\in\widehat{{\mathcal R}}_i+u\widehat{{\mathcal R}}_i$. Then}
$$\alpha(x)\mu(\beta(x))=\sum_{i=1}^r\theta_i(x)\left(a_i(x)\mu_i^{-1}(b_{\mu(i)}(x)\right)$$
\textit{where $\mu$ is the ring isomorphism from $\widehat{{\mathcal A}}+u\widehat{{\mathcal A}}$ onto ${\mathcal A}+u{\mathcal A}$ and $\mu_i^{-1}: \widehat{{\mathcal R}}_{\mu(i)}+u\widehat{{\mathcal R}}_{\mu(i)}\rightarrow {\mathcal R}_{i}+u{\mathcal R}_{i}$
being the inverse of $\mu_i$ defined in Lemma 4.6 for all $i=1,\ldots,r$}.

\vskip 3mm \noindent
  \textit{Proof}  Since $\mu$ is the ring isomorphism from $\widehat{{\mathcal A}}+u\widehat{{\mathcal A}}$ onto ${\mathcal A}+u{\mathcal A}$, by Lemma 4.5(iv) and Lemma 4.6 it follows that
\begin{eqnarray*}
\mu(\beta(x))&=&\sum_{j=1}^r\mu\left(\widehat{\theta}_j(x)b_j(x)\right)
   =\sum_{j=1}^r\theta_{\mu(j)}(x)\mu_{\mu(j)}^{-1}(b_j(x))
\end{eqnarray*}
where $\mu_{\mu(j)}^{-1}$ is the inverse of the ring isomorphism $\mu_{\mu(j)}$ from ${\mathcal R}_{\mu(j)}+u{\mathcal R}_{\mu(j)}$
onto $\widehat{{\mathcal R}}_j+u\widehat{{\mathcal R}}_j$. Hence $\mu_{\mu(j)}^{-1}(b_j(x))\in {\mathcal R}_{\mu(j)}+u{\mathcal R}_{\mu(j)}$
for all $j$. When $j=\mu(i)$, we have $i=\mu(j)$ by Lemma 4.5(i), and so $\mu_i^{-1}(b_{\mu(i)}(x))\in {\mathcal R}_i+u{\mathcal R}_i$,
which implies $a_i(x)\mu_i^{-1}(b_{\mu(i)}(x))\in {\mathcal R}_i+u{\mathcal R}_i$ for all $i$.

\par
   If $j\neq \mu(i)$, then $i\neq\mu(j)$ and hence $\theta_i(x)\theta_{\mu(j)}(x)=0$ by Theorem 2.9(i). From these, by
$\mu(\mu(i))=i$ and $\theta_i(x)^2=\theta_i(x)$ by Theorem 2.9(i) for all $i$, we deduce that
\begin{eqnarray*}
\alpha(x)\mu(\beta(x))&=&\left(\sum_{i=1}^r\theta_i(x)a_i(x)\right)\left(\sum_{j=1}^r\theta_{\mu(j)}(x)\mu_{\mu(j)}^{-1}(b_j(x))\right)\\
  &=&\sum_{i=1}^r\sum_{j=1}^r\left(\theta_i(x)a_i(x)\right)\left(\theta_{\mu(j)}(x)\mu_{\mu(j)}^{-1}(b_j(x))\right)\\
  &=&\sum_{i=1}^r\left(\theta_i(x)a_i(x)\right)\left(\theta_{\mu(\mu(i))}(x)\mu_{\mu(\mu(i))}^{-1}(b_j(x))\right),
\end{eqnarray*}
i.e. $\alpha(x)\mu(\beta(x))=\sum_{i=1}^r\theta_i(x)\left(a_i(x)\mu_i^{-1}(b_{\mu(i)}(x)\right)$.
\hfill $\Box$

\vskip 3mm \noindent
  {\bf Lemma 4.8} \textit{Using the notations of Lemma 4.6,
we have that
$\mu_i(f_i(x)^l)=\delta_i^lx^{-ld_i}f_{\mu(i)}(x)^l\in \widehat{{\mathcal R}}_{\mu(i)}$ for all $1\leq l\leq p^{k}s$}.

\vskip 3mm \noindent
  \textit{Proof} By Lemma 4.5(iii), we have $\widetilde{f}_i(x)=\delta_if_{\mu(i)}(x)$. From this and by
the definition of $\mu_i$ in Lemma 4.6, we deduce that
$
\mu_i(f_i(x)^l)=f_i(x^{-1})^l=x^{-ld_i}\left(x^{d_i}f_i(x^{-1})\right)^l
  =x^{-ld_i}\widetilde{f}_i(x)^l
  =\delta_i^lx^{-ld_i}f_{\mu(i)}(x)^l.
$
 \hfill $\Box$

\vskip 3mm \par
   Now, we give the dual code of each $(1+pw)$-constacyclic code over the ring $\mathbb{Z}_{p^s}+u\mathbb{Z}_{p^s}$
of length $N$ where $N=p^kn$.

\vskip 3mm \noindent
   {\bf Theorem 4.9} \textit{Let ${\mathcal C}$ be a $(1+pw)$-constacyclic code
over $\mathbb{Z}_{p^s}+u\mathbb{Z}_{p^s}$ of length $N$ with
${\mathcal C}=\oplus_{i=1}^r\theta_i(x)C_i$, where $C_i$ is an ideal of ${\mathcal R}_i+u{\mathcal R}_i$. Then
the dual code ${\mathcal C}^{\bot_E}$ is  a $(1+p\widehat{w})$-constacyclic code
over $\mathbb{Z}_{p^s}+u\mathbb{Z}_{p^s}$ of length $N$. Precisely, ${\mathcal C}^{\bot_E}$ is given by}
${\mathcal C}^{\bot_E}=\oplus_{j=1}^r\widehat{\theta}_j(x)D_j,$
\textit{where $D_j$ is an ideal of $\widehat{{\mathcal R}}_j+u\widehat{{\mathcal R}}_j$  given by the following table}:
{\footnotesize\begin{center}
\begin{tabular}{lll}\hline
case &  $C_i$  &  $D_{\mu(i)}$  \\ \hline
1.   & $\langle f_i(x)^{l_i}\rangle$ \ $(0\leq l_i\leq p^{k}s)$ &  $\langle f_{\mu(i)}(x)^{p^{k}s-l_i}\rangle$  \\
2.   &  $\langle u f_i(x)^{m_i}\rangle$ \ ($0\leq m_i\leq p^{k}s-1$) & $\langle f_{\mu(i)}(x)^{p^{k}s-m_i},u\rangle$ \\
3.   & $\langle f_i(x)^{l_i}+u f_i(x)^{t_i}h(x)\rangle$ & $\langle f_{\mu(i)}(x)^{p^{k}s-l_i}$ \\
     & ($h(x)\in \Delta_{l_i-t_i}^{(i)}$, $t_i\geq 2l_i-p^{k}s$, & $+u f_{\mu(i)}(x)^{p^{k}s+t_i-2l_i}\widehat{h}(x)\rangle$ \\
     & $0\leq t_i<l_i\leq p^{k}s-1$) & $\widehat{h}(x)=-\delta_i^{t_i-l_i}x^{(l_i-t_i)d_i} h(x^{-1})$\\
4.   & $\langle f_i(x)^{l_i}+uh(x)\rangle$  & $\langle f_{\mu(i)}(x)^{l_i}+u\widehat{h}(x)\rangle$ \\
     & $(h(x)\in \Delta_{p^{k}s-l_i}^{(i)}$, $\frac{1}{2}p^ks<l_i\leq p^{k}s-1$) & $\widehat{h}(x)=-\delta_i^{-l_i}x^{l_id_i}h(x^{-1})$ \\
5.   & $\langle f_i(x)^{l_i}+u f_i(x)^{t_i}h(x)\rangle$ & $\langle f_{\mu(i)}(x)^{l_i-t_i}+u \widehat{h}(x),u f_{\mu(i)}(x)^{p^{k}s-l_i}\rangle$  \\
     & $(h(x)\in \Delta_{p^{k}s-l_i}^{(i)}$, $t_i<2l_i-p^{k}s$,   & $\widehat{h}(x)=-\delta_i^{t_i-l_i}x^{(l_i-t_i)d_i} h(x^{-1})$ \\
     & $1\leq t_i<l_i\leq p^{k}s-1$)     & \\
6.   & $\langle f_i(x)^{l_i},u f_i(x)^{m_i}\rangle$ & $\langle f_{\mu(i)}(x)^{p^{k}s-m_i},u f_{\mu(i)}(x)^{p^{k}s-l_i}\rangle$ \\
     & $(0\leq m_i<l_i\leq p^{k}s-1)$ & \\
7.   & $\langle f_i(x)^{l_i}+uh(x), u f_i(x)^{m_i}\rangle$ & $\langle f_{\mu(i)}(x )^{p^ks-m_i}$ \\
     & $(h(x)\in \Delta_{m_i}^{(i)}$, $l_i+m_i\leq p^{k}s-1$,  & $+u f_{\mu(i)}(x)^{p^ks-l_i-m_i}\widehat{h}(x)\rangle$ \\
     & $1\leq m_i<l_i\leq p^{k}s-1$) & $\widehat{h}(x)=-\delta_i^{-l_i}x^{l_id_i}h(x^{-1})$\\
8. & $\langle f_i(x)^{l_i}+u f_i(x)^{t_i}h(x), u f_i(x)^{m_i}\rangle$ & $\langle u f_{\mu(i)}(x)^{p^{k}s+t_i-l_i-m_i}\widehat{h}(x)$ \\
   & $(h(x)\in \Delta_{m_i-t_i}^{(i)}$, $l_i+m_i\leq p^{k}s+t_i-1$, &  $ +f_{\mu(i)}(x)^{p^{k}s-m_i}, \ uf_{\mu(i)}(x)^{p^{k}s-l_i}\rangle $\\
   & $1\leq t_i<m_i<l_i\leq p^{k}s-1)$ & $\widehat{h}(x)=-\delta_i^{t_i-l_i}x^{(l_i-t_i)d_i}h(x^{-1})$ \\ \hline
\end{tabular}
\end{center}}

\vskip 3mm \noindent
  \textit{Proof} For each integer $i$, $1\leq i\leq r$, let
$J_i$ be an ideal of ${\mathcal R}_i+u{\mathcal R}_i$ given by one of the following eight cases:

\par
  (i) $J_i=\langle f_i(x)^{p^{k}s-l_i}\rangle$,  if $C_i=\langle f_i(x)^{l_i}\rangle$, where $0\leq l_i\leq p^{k}s$.

\par
  (ii) $J_i=\langle f_i(x)^{p^{k}s-m_i},u\rangle$,  if $C_i=\langle uf_i(x)^{m_i}\rangle$, where $0\leq m_i\leq p^{k}s-1$.

\par
  (iii)  $J_i=\langle f_i(x)^{p^{k}s-l_i}-u f_i(x)^{p^{k}s+t_i-2l_i}h(x)\rangle$, if $C_i=\langle f_i(x)^{l_i}+u f_i(x)^{t_i}h(x)\rangle$ where $h(x)\in \Delta_{l_i-t_i}^{(i)}$,
 $0\leq t_i<l_i\leq p^{k}s-1$ and $t_i\geq 2l_i-p^{k}s$.

\par
  (iv) $J_i=\langle f_i(x)^{l_i}-u h(x)\rangle$,  if $C_i=\langle f_i(x)^{l_i}+u h(x)\rangle$ where $h(x)\in \Delta_{p^ks-l_i}^{(i)}$,
$0<l_i\leq p^{k}s-1$ and $2l_i>p^{k}s$.

\par
  (v) $J_i=\langle f_i(x)^{l_i-t_i}-u h(x),u f_i(x)^{p^{k}s-l_i}\rangle$,  if $C_i=\langle f_i(x)^{l_i}+u f_i(x)^{t_i}h(x)\rangle$ where $h(x)\in \Delta_{p^{k}s-l_i}^{(i)}$,
$1\leq t_i<l_i\leq p^{k}s-1$ and $t_i<2l_i-p^{k}s$.

\par
  (vi) $J_i=\langle f_i(x)^{p^{k}s-m_i},u f_i(x)^{p^{k}s-l_i}\rangle$, if $C_i=\langle f_i(x)^{l_i},u f_i(x)^{m_i}\rangle$, where
$0\leq m_i<l_i\leq p^{k}s-1$.

\par
 (vii) $J_i=\langle f_i(x)^{p^{k}s-m_i}-u f_i(x)^{p^{k}s-l_i-m_i}h(x)\rangle$,  when $C_i=\langle f_i(x)^{l_i}+uh(x)$, $u f_i(x)^{m_i}\rangle$ in which $h(x)\in \Delta_{m_i}^{(i)}$,
$1\leq m_i<l_i\leq p^{k}s-1$ and $l_i+m_i\leq p^{k}s-1$.

\par
  (viii) $J_i=\langle f_i(x)^{p^{k}s-m_i}-u f_i(x)^{p^{k}s+t_i-l_i-m_i}h(x),u f_i(x)^{p^{k}s-l_i}\rangle$, if $C_i=\langle f_i(x)^{l_i}+u f_i(x)^{t_i}h(x),u f_i(x)^{m_i}\rangle$, where $h(x)\in \Delta_{m_i-t_i}^{(i)}$,
$1\leq t_i<m_i<l_i\leq p^{k}s-1$, $l_i+m_i\leq p^{k}s+t_i-1$.

\vskip 2mm \noindent
  Then it follows that
$C_i\cdot J_i=\{0\}$ by a straightforward computation.
Furthermore, by Theorem 3.3 we see that $|C_i||J_i|=p^{2p^{k}sd_i}$.

\par
  Let $1\leq i\leq r$ and denote $D_{\mu(i)}=\mu_i(J_i)$. By Lemma 4.6 we conclude that
$D_{\mu(i)}$ is an ideal of $\widehat{{\mathcal R}}_{\mu(i)}+u\widehat{{\mathcal R}}_{\mu(i)}$, $|D_{\mu(i)}|=|J_i|$.
and $\mu_i^{-1}(D_{\mu(i)})=J_i$. We set
${\mathcal D}=\oplus_{i=1}^r\widehat{\theta}_{\mu(i)}(x)D_{\mu(i)}=\sum_{j=1}^r\widehat{\theta}_{j}(x)D_{j}$
(mod $x^{p^kn}-(1+p\widehat{w})$). Then by the conclusion paralleling to Lemma 3.2, we see that
${\mathcal D}$ is a $(1+p\widehat{w})$-constacyclic code over $\mathbb{Z}_{p^s}+u\mathbb{Z}_{p^s}$ of length $p^kn$. Moreover,
by Lemma 4.7 and $C_i\cdot\mu_i^{-1}(D_{\mu(i)})=C_i\cdot J_i=\{0\}$ it follows that
$${\mathcal C}\cdot \mu({\mathcal D})=(\sum_{i=1}^r\theta_i(x)C_i)\cdot\mu(\sum_{j=1}^r\widehat{\theta}_{j}(x)D_{j})
=\sum_{i=1}^r\theta_i(x)\left(C_i\cdot\mu_i^{-1}(D_{\mu(i)})\right)=\{0\},$$
which implies ${\mathcal D}\subseteq {\mathcal C}^{\bot_E}$ by Lemma 4.4. On the other hand, by Lemma 3.1 and Corollary 4.3(iii)
we have
\begin{eqnarray*}
|{\mathcal C}||{\mathcal D}|&=&\left(\prod_{i=1}^r|\theta_i(x)C_i|\right)\left(\prod_{i=1}^r|\widehat{\theta}_{\mu(i)}(x)D_{\mu(i)}|\right)
 =\prod_{i=1}^r\left(|C_i||D_{\mu(i)}|\right)\\
  &=&\prod_{i=1}^r\left(|C_i||J_i|\right)=\prod_{i=1}^rp^{2p^ksd_i}
  =p^{2p^ks\sum_{i=1}^rd_i}=(p^{2s})^{p^kn}\\
  &=&|\mathbb{Z}_{p^s}+u \mathbb{Z}_{p^s}|^{p^kn}.
\end{eqnarray*}
Since both ${\mathcal C}$ and ${\mathcal D}$ are linear codes over $\mathbb{Z}_{p^s}+u \mathbb{Z}_{p^s}$ of length $p^kn$
and $\mathbb{Z}_{p^s}+u \mathbb{Z}_{p^s}$ is a finite Frobenius ring, from the theory of linear codes over Frobenius rings (see [13], for example)
we deduce that ${\mathcal C}^{\bot_E}={\mathcal D}$.

\par
  Finally, we give the precise representation of each $D_{\mu(i)}$, $1\leq i\leq r$.

\par
  Case 1. Let $J_i=\langle f_i(x)^{p^{k}s-l_i}\rangle$ as in (i).
Since $x$ is an invertible element of $\widehat{{\mathcal R}}_{\mu(i)}+u\widehat{{\mathcal R}}_{\mu(i)}$, by Lemmas 4.6 and 4.8 we have
\begin{eqnarray*}
D_{\mu(i)}&=&\langle\mu_i(f_i(x)^{p^{k}s-l_i})\rangle
=\langle\delta_i^{p^{k}s-l_i}x^{-(p^{k}s-l_i)d_i}f_{\mu(i)}(x)^{p^{k}s-l_i}\rangle\\
&=&\langle f_{\mu(i)}(x)^{p^{k}s-l_i}\rangle.
\end{eqnarray*}

\par
  Cases 2 and 6 can be obtained similarly as Case 1.

\par
  Case 3. Let $J_i=\langle f_i(x)^{p^{k}s-l_i}-u f_i(x)^{p^{k}s+t_i-2l_i}h(x)\rangle$ as in (iii). By Lemmas 4.6 and 4.8 it follows that
\begin{eqnarray*}
D_{\mu(i)}&=&\langle\mu_i(f_i(x)^{p^{k}s-l_i})-u\cdot \mu_i(f_i(x)^{p^{k}s+t_i-2l_i})\mu_i(h(x))\rangle\\
 &=&\langle \delta_i^{p^{k}s-l_i}x^{-({p^{k}s-l_i})d_i}f_{\mu(i)}(x)^{p^{k}s-l_i}\\
 &&-u(\delta_i^{p^{k}s+t_i-2l_i}x^{-(p^{k}s+t_i-2l_i)d_i}f_{\mu(i)}(x)^{p^{k}s+t_i-2l_i})h(x^{-1})\rangle\\
 &=&\langle f_{\mu(i)}(x)^{2^{k+1}-l_i}+uf_{\mu(i)}(x)^{2^{k+1}+t_i-2l_i}\widehat{h}(x)\rangle,
\end{eqnarray*}
where $\widehat{h}(x)=-\delta_i^{t_i-l_i}x^{(l_i-t_i)d_i}h(x^{-1})\in \widehat{{\mathcal A}}_{\mu(i)}^{\times}$.

\par
  Case 4. Let $J_i=\langle f_i(x)^{l_i}-uw(x)\rangle$ as in (iv). By Lemmas 4.6 and 4.8 we have
\begin{eqnarray*}
D_{\mu(i)}&=&\langle \mu_i(f_i(x)^{l_i})-u\mu_i( (h(x))\rangle
 =\langle \delta_i^{l_i}x^{-l_id_i}f_{\mu(i)}(x)^{l_i}-u  h(x^{-1})\rangle\\
&=&\langle f_{\mu(i)}(x)^{l_i}+u\widehat{h}(x)\rangle,
\end{eqnarray*}
where $\widehat{h}(x)=-\delta_i^{-l_i}x^{l_id_i}h(x^{-1})\in \widehat{{\mathcal A}}_{\mu(i)}^{\times}$.

\par
  Cases 5, 7 and 8 can be obtained similarly as Cases 1, 3 and 4.
\hfill $\Box$

\vskip 3mm \par
   Finally, we consider the self-duality of the constacyclic codes over $\mathbb{Z}_{p^s}+u\mathbb{Z}_{p^s}$.
When $p$ is odd, by Lemma 2.6 we know that $(1+pw)^{-1}\neq 1+pw$ for all $w\in \mathbb{Z}_{p^s}^{\times}$.
Let $p=2$ and $w\in \mathbb{Z}_{2^s}^{\times}$. Then it is clear that $(1+2w)^{-1}=1+2w$ if and only if
$w=2^{s-2}-1$ or $2^{s-1}-1$ when $s\geq 3$, and $w=1$ when $s=2$.

\vskip 3mm \par
   Now, let $p=2$, and $w=2^{s-2}-1$ or $2^{s-1}-1$ if $s\geq 3$, and $w=1$ if $s=2$. We assume $N=2^kn$ where $n$ is an odd positive
integer.
Then $(1+2w)^2=1\in \mathbb{Z}_{2^s}^{\times}$. In this case, Using the notations of Lemma 2.6 and Lemma 4.1 we have
$1+2w_0=1+2w=1+2\widehat{w}=1+2\widehat{w}_0$. Furthermore, using Equations (2.5)
and (4.1) it follows that
\begin{equation}
\theta_i(x)=\widehat{\theta}_i(x)\equiv a_i((1+2w)x^{2^k})F_i((1+2w)x^{2^k})
\ ({\rm mod} \ x^{2^kn}-(1+2w)),
\end{equation}
which implies ${\mathcal A}_i=\widehat{{\mathcal A}}_i$ for all $i=1,\ldots,r$. Then
as a corollary of Lemma 3.2, Theorems 3.3 and 4.9, we can present all distinct self-dual $(1+2w)$-constacyclic
codes over the ring $\mathbb{Z}_{2^s}+u\mathbb{Z}_{2^s}$ by the following theorem.

\vskip 3mm \noindent
  {\bf Theorem 4.10} \textit{Let $w=2^{s-2}-1$ or $2^{s-1}-1$ when $s\geq 3$, and $w=1$ when $s=2$.
Using the notations in Theorem 4.9, Lemma 4.5$({\rm ii})$ and Equation $(12)$,
let ${\mathcal C}$ be a $(1+2w)$-constacyclic code
over $\mathbb{Z}_{2^s}+u\mathbb{Z}_{2^s}$ of length $N$ with
${\mathcal C}=\oplus_{i=1}^r\theta_i(x)C_i$, where $C_i$ is an ideal of ${\mathcal R}_i+u{\mathcal R}_i$. Then
${\mathcal C}$ is self-dual if and only if for each integer $i$, $1\leq i\leq r$, $C_i$ satisfies one
of the following conditions}:

\vskip 2mm \par
  $\bullet$ \textit{If $i=\lambda+j$ where $1\leq j\leq \epsilon$, $(C_{i},C_{i+\epsilon})$ are given by the following table}:
{\footnotesize \begin{center}
\begin{tabular}{lll}\hline
case &  $C_i$  &  $C_{i+\epsilon}$  \\ \hline
1.   & $\langle f_i(x)^{l_i}\rangle$ \ $(0\leq l_i\leq 2^{k}s)$ &  $\langle f_{i+\epsilon}(x)^{2^{k}s-l_i}\rangle$  \\
2.   &  $\langle u f_i(x)^{m_i}\rangle$ \ ($0\leq m_i\leq 2^{k}s-1$) & $\langle f_{i+\epsilon}(x)^{2^{k}s-m_i},u\rangle$ \\
3.   & $\langle f_i(x)^{l_i}+u f_i(x)^{t_i}h(x)\rangle$ & $\langle f_{i+\epsilon}(x)^{2^{k}s-l_i}$ \\
     & ($w(x)\in \Delta_{l_i-t_i}^{(i)}$, $t_i\geq 2l_i-2^{k}s$, & $+u f_{i+\epsilon}(x)^{2^{k}s+t_i-2l_i}\widehat{h}(x)\rangle$ \\
     & $0\leq t_i<l_i\leq 2^{k}s-1$) & $\widehat{h}(x)=-\delta_i^{t_i-l_i}x^{(l_i-t_i)d_i} h(x^{-1})$\\
4.   & $\langle f_i(x)^{l_i}+uh(x)\rangle$  & $\langle f_{i+\epsilon}(x)^{l_i}+u\widehat{h}(x)\rangle$ \\
     & $(h(x)\in \Delta_{2^{k}s-l_i}^{(i)}$, & $\widehat{h}(x)=-\delta_i^{-l_i}x^{l_id_i}h(x^{-1})$ \\
     &  $2^{k-1}s+1\leq l_i\leq 2^{k}s-1$) & \\
5.   & $\langle f_i(x)^{l_i}+u f_i(x)^{t_i}h(x)\rangle$ & $\langle f_{i+\epsilon}(x)^{l_i-t_i}+u\widehat{h}(x),u f_{i+\epsilon}(x)^{2^{k}s-l_i}\rangle$  \\
     & $(h(x)\in \Delta_{2^{k}s-l_i}^{(i)}$, $t_i<2l_i-2^{k}s$,   & $\widehat{h}(x)=-\delta_i^{t_i-l_i}x^{(l_i-t_i)d_i} h(x^{-1})$ \\
     & $1\leq t_i<l_i\leq 2^{k}s-1$)     & \\
6.   & $\langle f_i(x)^{l_i},u f_i(x)^{m_i}\rangle$ & $\langle f_{i+\epsilon}(x)^{2^{k}s-m_i},u f_{i+\epsilon}(x)^{2^{k}s-l_i}\rangle$ \\
     & $(0\leq m_i<l_i\leq 2^{k}s-1)$ & \\
7.   & $\langle f_i(x)^{l_i}+uh(x), u f_i(x)^{m_i}\rangle$ & $\langle f_{i+\epsilon}(x)^{2^{k}s-m_i}$ \\
     & $(h(x)\in \Delta_{m_i}^{(i)}$, $l_i+m_i\leq 2^{k}s-1$,  & $+u f_{i+\epsilon}(x)^{2^{k}s-l_i-m_i}\widehat{h}(x)\rangle$ \\
     & $1\leq m_i<l_i\leq 2^{k}s-1$) & $\widehat{h}(x)=-\delta_i^{-l_i}x^{l_id_i}h(x^{-1})$  \\
8. & $\langle f_i(x)^{l_i}+u f_i(x)^{t_i}h(x), u f_i(x)^{m_i}\rangle$ & $\langle u f_{i+\epsilon}(x)^{2^{k}s+t_i-l_i-m_i}\widehat{h}(x)$ \\
   & $(h(x)\in \Delta_{m_i-t_i}^{(i)}$,  $l_i+m_i\leq 2^{k}s+t_i-1$,  &  $+f_{i+\epsilon}(x)^{2^{k}s-m_i}, \ u f_{i+\epsilon}(x)^{2^{k}s-l_i}\rangle$\\
   & $1\leq t_i<m_i<l_i\leq 2^{k}s-1)$ & $\widehat{h}(x)=-\delta_i^{t_i-l_i}x^{(l_i-t_i)d_i}h(x^{-1})$ \\ \hline
\end{tabular}
\end{center}}

\vskip 2mm \par
  $\bullet$ \textit{If $1\leq i\leq \lambda$, $C_i$ is given by one of the following six cases}:

\vskip 2mm \par
  (i) \textit{$C_i=\langle f_i(x)^{2^{k-1}s}\rangle$}. \
  (ii) \textit{$C_i=\langle u \rangle$}.

\vskip 2mm \par
  (iii) \textit{$C_i=\langle f_i(x)^{2^{k-1}s}+u f_i(x)^{t_i}h(x)\rangle$, where
$0\leq t_i\leq 2^{k-1}s-1$ and $h(x)\in \triangle_{2^{k-1}s-t_i}^{(i)}$ satisfying
$h(x)+\delta_i^{t_i-2^{k-1}s}x^{(2^{k-1}s-t_i)d_i}h(x^{-1})\equiv 0$ $($mod $f_i(x)^{2^{k-1}s-t_i}$$)$}.

\vskip 2mm \par
  (iv) \textit{$C_i=\langle f_i(x)^{l_i}+u h(x)\rangle$, where
$2^{k-1}s+1\leq l_i\leq 2^{k}s-1$ and $h(x)\in \Delta_{2^{k}s-l_i}^{(i)}$ satisfying
$h(x)+\delta_i^{-l_i}x^{l_id_i}h(x^{-1})\equiv 0$ $($mod $f_i(x)^{2^{k}s-l_i}$$)$}.

\vskip 2mm \par
  (v) \textit{$C_i=\langle f_i(x)^{l_i},u f_i(x)^{2^{k}s-l_i}\rangle$, where
$2^{k-1}s+1<l_i\leq 2^{k}s-1$}.

\vskip 2mm \par
  (vi) \textit{$C_i=\langle f_i(x)^{l_i}+u f_i(x)^{t_i}w(x), u f_i(x)^{2^{k}s-l_i}\rangle$, where
$1\leq t_i<2^{k}s-l_i$, $2^{k-1}s+1\leq l_i\leq 2^{k}s-1$ and $h(x)\in \Delta_{2^{k}s-l_i-t_i}^{(i)}$ satisfying}
\begin{center}
$h(x)+\delta_i^{t_i-l_i}x^{(l_i-t_i)d_i}h(x^{-1})\equiv 0$ (mod $f_i(x)^{2^{k}s-l_i-t_i}$).
\end{center}

%%%%%%%%%%%%%%%%%%%%%%%%%%%%%%%%%%%%%%%%%%%%%%%%%%%%%%%%%%%%%%%%%%%%%%%%

%%%%%%%%%%%%%%%%%%%%%%%%%%%%%%%%%%%%%%%%%%%%%%%%%%%%%%%%%%%%%%%%%%%%%%%%%%

\section{\bf An Example}
 In this section, we consider how to construct all distinct self-dual $3$-constacyclic codes over $\mathbb{Z}_{8}+u\mathbb{Z}_{8}$ ($u^2=0$) of length $14$.
In this case, we have $N=2^kn$ with $k=1$, $n=7$, $p=2$, $s=3$ and $1+pw=3$ satisfying $3^2=1$ in $\mathbb{Z}_{8}$.
It is known that
$y^7-1=f_1(y)f_2(y)f_3(y),$
where
$f_1(y)=y-1$, $f_2(y)=y^3+6y^2+5y+7$ and $f_3(y)=y^3+3y^2+2y+7$
are pairwise coprime minic basic irreducible polynomials in $\mathbb{Z}_{8}[y]$.

\par
   As $d_1=1$, $d_2=d_3=3$ and $r=3$, by Theorem 3.3 and Corollary 3.4, the number $N$ of $3$-constacyclic codes over $\mathbb{Z}_8+u\mathbb{Z}_8$ of length $14$
is equal to $N=\prod_{i=1}^3(2^{3d_i}+5\cdot 2^{2d_i}+9\cdot 2^{d_i}+13)=59\cdot 917^2
=49,612,451$.

Obviously, $\widetilde{f}_1(y)=\delta_1f_1(y)$ and $\widetilde{f}_2(y)=\delta_2f_3(y)$ where $\delta_1=\delta_2=-1$, which implies that
$\mu(1)=1$ and $\mu(2)=3$. Hence $\lambda=\epsilon=1$.

\par
  For each integer $i$, $1\leq i\leq 3$, we denote
$F_i(y)=\frac{y^7-1}{f_i(y)}$, and find polynomials $a_i(y),b_i(y)\in \mathbb{Z}_{8}[y]$
satisfying $a_i(y)F_i(y)+b_i(y)f_i(y)=1$. Then set $\varrho_i(y)\equiv a_i(y)F_i(y)$ (mod $y^7-1$). By Equation (4.2) in Section 4, it follows that
$\theta_i(x)\equiv\varrho_i(3x^2)$ (mod $x^{14}+1$) in $\mathbb{Z}_{8}[x]$. Precisely, we have

\vskip 2mm\par
   $\theta_1(x)=7+5x^2+7x^4+5x^6+7x^8+5x^{10}+7x^{12}$;

\par
   $\theta_2(x)=5+x^2+3x^4+2x^6+3x^8+2x^{10}+6x^{12}$;

\par
  $\theta_3(x)=5+2x^2+6x^4+x^6+6^8+x^{10}+3x^{12}$.

\vskip 2mm\noindent
  Using the notations in Section 3, for each integer $j$, $1\leq j\leq 4$ we have that ${\mathcal R}_i=\mathbb{Z}_{8}[x]/\langle f_i(3x^2)\rangle$,
which is a finite chain ring with the maximal ideal $f_i(x)$ and the nilpotency index of $f_i(x)$ ie equal to $2\cdot 3=6$.

\par
  Now, by Theorem 4.10 all distinct self-dual $3$-constacyclic codes over $\mathbb{Z}_{8}+u\mathbb{Z}_{8}$ ($u^2=0$) of length $14$
are given by
\begin{center}
${\mathcal C}=\theta_1(x)C_1\oplus \theta_2(x)C_2\oplus \theta_3(x)C_3=\sum_{i=1}^3\theta_i(x)C_i$ (mod $x^{14}-3$),
\end{center}
where $C_i$ is an ideal of the ring ${\mathcal R}_i+u{\mathcal R}_i$ satisfying one of the following conditions:

\vskip 2mm \par
  $\bullet$ $C_1$ is one of the following seven ideals:

\vskip 2mm \par
   $\langle (x-1)^{3}\rangle$, $\langle u\rangle$, $\langle (x-1)^{3}+u(x-1)\rangle$, $\langle(x-1)^{5}+u\rangle$,
   $\langle(x-1)^{4},u(x-1)^2\rangle$,

\vskip 2mm \par
   $\langle(x-1)^{5},u(x-1)\rangle$, $\langle(x-1)^{4}+u(x-1),u(x-1)^2\rangle$.

\vskip 2mm \par
  $\bullet$ $(C_2,C_3)$ is given by the following table, where
$C_i$ is an ideal of ${\mathcal R}_i+u{\mathcal R}_i$ for $i=2,3$, and $L_{(C_2,C_3)}$ is the
number of pairs $(C_2,C_3)$ on the same line.
{\small \begin{center}
\begin{tabular}{lll}\hline
case &  $C_2$  &  $C_{3}$  \\ \hline
1.   & $\langle f_2(x)^{l}\rangle$ \ $(0\leq l\leq 6)$ &  $\langle f_{3}(x)^{6-l}\rangle$  \\
2.   &  $\langle u f_2(x)^{m}\rangle$ \ ($0\leq m\leq 5$) & $\langle f_{3}(x)^{6-m},u\rangle$ \\
3.   & $\langle f_2(x)^{l}+u f_2(x)^{t}h(x)\rangle$ & $\langle f_{3}(x)^{6-l}+u f_{3}(x)^{6+t-2l}\widehat{h}(x)\rangle$ \\
     & ($h(x)\in \Delta_{l-t}^{(2)}$, $t\geq 2l-6$, & $\widehat{h}(x)=(-1)^{l-t+1}x^{3(l-t)} h(x^{-1})$ \\
     & $0\leq t<l\leq 5$) & \\
4.   & $\langle f_2(x)^{l}+uh(x)\rangle$  & $\langle f_{3}(x)^{l}+u\widehat{h}(x)\rangle$ \\
     & $(h(x)\in \Delta_{6-l}^{(2)},l=4,5$) & $\widehat{h}(x)=(-1)^{l+1}x^{3l}h(x^{-1})$ \\
5.   & $\langle f_2(x)^{l}+u f_2(x)^{t}h(x)\rangle$ & $\langle f_{3}(x)^{l-t}+u\widehat{h}(x),u f_{3}(x)^{6-l}\rangle$  \\
     & $(h(x)\in \Delta_{6-l}^{(2)}$,    & $\widehat{h}(x)=(-1)^{l-t+1}x^{3(l-t)d_i} h(x^{-1})$ \\
     & $(m,l)\in\{(1,4),(1,5),(2,5),(3,5)\}$)     & \\
6.   & $\langle f_2(x)^{l},u f_2(x)^{m}\rangle$ & $\langle f_{3}(x)^{6-m},u f_{3}(x)^{6-l}\rangle$ \\
     & $(0\leq m<l\leq 5)$ & \\
7.   & $\langle f_2(x)^{l}+uh(x), u f_2(x)^{m}\rangle$ & $\langle f_{3}(x)^{6-m}+u f_{3}(x)^{6-l-m}\widehat{h}(x)\rangle$ \\
     & $(h(x)\in \Delta_{m}^{(2)}$,  & $\widehat{h}(x)=(-1)^{l+1}x^{3l}h(x^{-1})$ \\
     & $(m,l)\in\{(1,2),(1,3),(1,4),(2,3)\})$ &   \\
8. & $\langle f_2(x)^{l}+u f_2(x)^{t}h(x), u f_2(x)^{m}\rangle$ & $\langle f_{3}(x)^{6-m}+u f_{3}(x)^{6+t-l-m}\widehat{h}(x),$ \\
   & $(h(x)\in \Delta_{1}^{(2)}$, \ $(t,m,l)\in\{(1,2,3),$  &  $u f_{3}(x)^{6-l}\rangle$\\
   &  $(1,2,4),(2,3,4)\})$  & $\widehat{h}(x)=(-1)^{l-t+1}x^{3(l-t)}h(x^{-1})$\\
 \hline
\end{tabular}
\end{center}}

\noindent
where $\Delta^{(2)}_k=\{\sum_{j=0}^{k-1}b_j(x)f_2(x)^j\mid b_j(x)\in {\mathcal T}_2, b_0(x)\neq 0, 0\leq j\leq k-1\}$ with ${\mathcal T}_2=\{a_0+a_1x+a_2x^2\mid a_0,a_1,a_2\in \{0,1\}\}$, for $k=1,2,3$.

\vskip 2mm \par
    Therefore, the number of self-dual $3$-constacyclic codes over $\mathbb{Z}_8+u\mathbb{Z}_8$ of length
$14$ is equal to $7\cdot 917=6419$.

%%%
%%%

%%%
%%% Acknowledgments
%%%
\section*{\bf Acknowledgments}
 Part of this work was done when Yonglin Cao was visiting Chen Institute of Mathematics, Nankai University, Tianjin, China. Yonglin Cao would like to thank the institution for the kind hospitality. This research is
supported in part by the National Natural Science Foundation of
China (Grant Nos. 11671235, 11471255).

%%%
%%% Appendices
%%%
\appendix
\section{A direct proof for Equation (2.1)}
 \vskip 3mm \noindent
  Denote $\eta=1+pw_0\in \mathbb{Z}_{p^s}^\times$. By Lemma 2.4(ii), we have
$f(\eta^{-1} x^{p^k})=\prod_{i=0}^{d-1}(\eta^{-1}x^{p^k}-\zeta^{p^i})=\eta^{-d}\prod_{i=0}^{d-1}(x^{p^k}-\eta\zeta^{p^i})$.
As $x^{p^k}-\eta\zeta^{p^i}=(x^{p^k}-\zeta^{p^i})-pw_0\zeta^{p^i}$,
we have $f(\eta^{-1} x^{p^k})=\eta^{-d}\prod_{i=0}^{d-1}(x^{p^k}-\zeta^{p^i})-pg_1(x)+p^2g_2(x)$ where
\begin{equation}
g_1(x)=\eta^{-d}w_0\sum_{i=0}^{d-1}\zeta^{p^i}\prod_{0\leq j\neq i\leq d-1}(x^{p^k}-\zeta^{p^j})
\end{equation}
and $g_2(x)\in \Gamma[x]$. By Lemma 2.4(ii) we have
\begin{eqnarray*}
f(x)^{p^k}&=&\prod_{i=0}^{d-1}(x-\zeta^{p^i})^{p^k}=\prod_{i=0}^{d-1}\left((x^{p^k}+(-\zeta^{p^i})^{p^k})+pa_i(x)\right)\\
&=&\prod_{i=0}^{d-1}\left((x^{p^k}-\zeta^{p^{i+k}})+pa_i(x)\right)\\
&=&\prod_{i=0}^{d-1}(x^{p^k}-\zeta^{p^{i+k}})+ph_1(x)+p^2h_2(x),
\end{eqnarray*}
where $a_i(x)\in \Gamma[x]$ satisfies $(x-\zeta^{p^i})^{p^k}=x^{p^k}-(\zeta^{p^i})^{p^k}+pa_i(x)$,
\begin{equation}
h_1(x)=\sum_{i=0}^{d-1}a_i(x)\prod_{0\leq j\neq i\leq d-1}(x^{p^k}-\zeta^{p^{j+k}})
\end{equation}
and $h_2(x)\in \Gamma[x]$. In fact, we have
$a_i(x)=\sum_{t=1}^{p^k-1}\frac{1}{p}\left(\begin{array}{c}p^k\cr t\end{array}\right)(-\zeta^{p^i})^tx^{p^k(p^k-t)}$
where
$\frac{1}{p}\left(\begin{array}{c}p^k\cr t\end{array}\right)=\frac{1}{p}\frac{p^k!}{(p^k-t)!t!}\in \mathbb{Z}$ for
all $t=1,\ldots,p^k-1$, when $p$ is odd. Let $p=2$. Then
$a_i(x)=\zeta^{2^{i+k}}+\sum_{t=1}^{2^k-1}\frac{1}{2}\left(\begin{array}{c}2^k\cr t\end{array}\right)(-\zeta^{2^i})^tx^{2^k(2^k-t)}$
where
$\frac{1}{2}\left(\begin{array}{c}2^k\cr t\end{array}\right)=\frac{1}{2}\frac{2^k!}{(2^k-t)!t!}\in \mathbb{Z}$ for
all $t=1,\ldots,2^k-1$.

\par
   By Lemma 2.4(i), we know that $\zeta^{p^{d}}=\zeta=\zeta^{p^0}$, which implies
$\prod_{i=0}^{d-1}(x^{p^k}-\zeta^{p^i})=\prod_{i=0}^{d-1}(x^{p^k}-\zeta^{p^{i+k}})$, and hence
$$\eta^df(\eta^{-1}x^{p^k})=\prod_{i=0}^{d-1}(x^{p^k}-\zeta^{p^{i+k}})-\eta^d\left(pg_1(x)-p^2g_2(x)\right).$$
Now, denote $\vartheta(x)=h_1(x)+\eta^dg_1(x)+p(h_2(x)-\eta^dg_2(x))$. Then
\begin{equation}
f(x)^{p^k}=\eta^df(\eta^{-1}x^{p^k})+p\vartheta(x),
\end{equation}
which implies $\vartheta(x)\in \mathbb{Z}_{p^s}[x]$, hence we see that
$$\vartheta(x)=\frac{1}{p}\left(f(x)^{p^k}-\eta^df(\eta^{-1}x^{p^k})\right)\in \mathbb{Z}[x]$$
by $(1+pw_0)^{-1}\in \mathbb{Z}_{p^s}=\{0,1,\ldots,p^s-1\}$. Moreover, we have
\begin{eqnarray*}
\overline{\vartheta}(x)&=& h_1(x)+\eta^dg_1(x)+p(h_2(x)-\eta^dg_2(x)) \ ({\rm mod} \ p)\\
 &=& \overline{h}_1(x)+\overline{g}_1(x).
\end{eqnarray*}

\par
   By Lemma 2.4(i), we know that $\zeta^{p^{d}}=\zeta=\zeta^{p^0}$ in the Galois ring $\Gamma$, which implies
$\overline{\zeta}^{p^{d}}=\overline{\zeta}^{p^0}$ in the finite field $\overline{\Gamma}$.
From this and by Equations (A.1) and (A.2), we have
\begin{eqnarray*}
\overline{g}_1(x)&=&\overline{w}_0\sum_{i=0}^{d-1}\overline{\zeta}^{p^i}\prod_{0\leq j\neq i\leq d-1}(x^{p^k}-\overline{\zeta}^{p^j})\\
  &=&\overline{w}_0\sum_{i=0}^{d-1}\overline{\zeta}^{p^{i+k}}\prod_{0\leq j\neq i\leq d-1}(x^{p^k}-\overline{\zeta}^{p^{j+k}})\\
  &=&\overline{w}_0\sum_{i=0}^{d-1}\overline{\zeta}^{p^{i+k}}\prod_{0\leq j\neq i\leq d-1}(x-\overline{\zeta}^{p^{j}})^{p^k},
\end{eqnarray*}
$$\overline{h}_1(x)=\sum_{i=0}^{d-1}\overline{a}_i(x)\prod_{0\leq j\neq i\leq d-1}(x^{p^k}-\overline{\zeta}^{p^{j+k}})
=\sum_{i=0}^{d-1}\overline{a}_i(x)\prod_{0\leq j\neq i\leq d-1}(x-\overline{\zeta}^{p^{j}})^{p^k}.$$
By Lemma 2.4(i), we know that $\overline{\zeta},\overline{\zeta}^p,\ldots,\overline{\zeta}^{p^{d-1}}$ are all distinct roots
of $\overline{f}(x)$ in $\overline{\Gamma}$. Then for any integer $i$, $0\leq i\leq d-1$, we have
$\prod_{0\leq j\leq d-1, j\neq i}(\overline{\zeta}^{p^{i}}-\overline{\zeta}^{p^{j}})^{p^k}=\left(\prod_{0\leq j\leq d-1, j\neq i}(\overline{\zeta}^{p^{i}}-\overline{\zeta}^{p^{j}})\right)^{p^k}\neq 0$
and $\prod_{0\leq j\leq d-1, j\neq i}(\overline{\zeta}^{p^{t}}-\overline{\zeta}^{p^{j}})^{p^k}=0$ for all $t\neq i$.
Therefore,
$$\overline{g}_1(\overline{\zeta}^{p^{i}})
=\overline{w}_0{\zeta}^{p^{i+k}}\prod_{0\leq j\leq d-1,j\neq i}(\overline{\zeta}^{p^{i}}-\overline{\zeta}^{p^{j}})^{p^k}\neq 0.$$
By $(x-\zeta^{p^i})^{p^k}=x^{p^k}-(\zeta^{p^i})^{p^k}+pa_i(x)$ in $\Gamma[x]$, we have
$$pa_i(\zeta^{p^i})=(\zeta^{p^i}-\zeta^{p^i})^{p^k}-((\zeta^{p^i})^{p^k}-(\zeta^{p^i})^{p^k})=0 \ {\rm in} \ \Gamma.$$
Since $\Gamma$ is a finite chain with the unique maximal ideal generated by $p$ and $s\geq 2$ is the nilpotency index of $p$, we conclude
that $a_i(\zeta^{p^i})=p^{s-1}\alpha$ for some $\alpha\in \Gamma$ (see [14], for example), which then implies
$\overline{a}_i(\overline{\zeta}^{p^i})=\overline{a_i(\zeta^{p^i})}=0$ and hence $\overline{h}_1(\overline{\zeta}^{p^i})=0$ in $\overline{\Gamma}$.

\par
   As state above, we conclude that $\overline{\vartheta}(\overline{\zeta}^{p^i})=\overline{g}_1(\overline{\zeta}^{p^{i}})\neq 0$ for all $i=0,1,\ldots,d-1$. From this and by
$\overline{f(\eta^{-1}x^{p^k})}=\overline{f}(x^{p^k})=\overline{f}(x)^{p^k}=\prod_{i=0}^{d-1}(x-\overline{\zeta}^{p^i})^{p^k}$,
we deduce that $\overline{\vartheta}(x)$ and $\overline{f(\eta^{-1}x^{p^k})}$ are coprime polynomials in $\mathbb{F}_p[x]$. Hence
$\vartheta(x)$ and $f(\eta^{-1}x^{p^k})$ are coprime polynomials in $\mathbb{Z}_{p^s}[x]$ by [16] Lemma 13.5.
 So there exist $a(x),b(x)\in \mathbb{Z}_{p^s}[x]$ such that
$a(x)\vartheta(x)+b(x)f(\eta^{-1}x^{p^k})=1,$
which implies that $\vartheta(x)$ is an invertible element of ${\mathcal R}=\mathbb{Z}_{p^s}[x]/\langle f(\eta^{-1}x^{p^k})\rangle$. By Equation (A.3), we have
$f(x)^{p^k}=p\vartheta(x)$ in ${\mathcal R}$ where $\vartheta(x)\in {\mathcal R}^{\times}$.

%------------------------------------------------------------------------------------%

\end{document}